\documentclass[5p,times]{elsarticle}

\journal{Chemical Physics}

\usepackage{amsmath,amssymb,amsfonts}

\usepackage{bm}

\usepackage{color,colordvi,graphicx}
\usepackage{epstopdf}

\newcommand{\mtrx}[1]{\ensuremath{\bm{\mathsf{#1}}}}

\definecolor{mybeige}{RGB}{220, 216, 197}

\begin{document}

\begin{frontmatter}

\title{Jost function description of near threshold resonances for
  coupled-channel scattering}

\author{I. Simbotin}


\author{R. C\^ot\'e\corref{robin}}
\cortext[robin]{Corresponding author}
\ead{rcote@phys.uconn.edu}

\address{Department of Physics, University of Connecticut,
  2152 Hillside Rd., Storrs, CT 06269, USA}



\date{\today}

\begin{abstract}
  We study the effect of resonances near the threshold of low energy
  ($\varepsilon$) reactive scattering processes, and find an anomalous
  behavior of the $s$-wave cross sections.  For reaction and inelastic
  processes, the cross section exhibits the energy dependence
  $\sigma\sim\varepsilon^{-3/2}$ instead of the standard Wigner's law
  threshold behavior $\sigma\sim\varepsilon^{-1/2}$. Wigner's law is
  still valid as $\varepsilon\rightarrow 0$, but in a narrow range of
  energies. We illustrate these effects with two reactive systems, a
  low-reactive system (H$_2$ + Cl) and a more reactive one (H$_2$ +
  F). We provide analytical expressions, and explain this anomalous
  behavior using the properties of the Jost functions.  We also
  discuss the implication of the reaction rate coefficients behaving
  as $K\sim 1/T$ at low temperatures, instead of the expected constant
  rate of the Wigner regime in ultracold physics and chemistry.
\end{abstract}


\end{frontmatter}


\section{Introduction}

Ultracold gases allow a high level of control over the interaction by
using Feshbach resonances \cite{RMP-FR} or by orienting ultracold
molecules \cite{paper-JILA,Jason-PRL}. In addition to the study of
phenomena in degenerate quantum gases ranging from the BEC-BCS
cross-over regime to solitons and multi-component condensates
\cite{RMP-bose,RMP-fermi}, such control also permits the investigation
of exotic three-body Efimov states \cite{efimov}.  In parallel, the
rapid advances in forming cold molecules
\cite{carr2009,dulieu2011,mol-papers,Mol-RC,Pellegrini} have made
possible studies of cold chemical reactions
\cite{paper-JILA,sawyer2011} and their control \cite{quemener2012},
with application to a growing range of fields, such as quantum
information processing \cite{DeMille-QC,Lena-1,Lena-2}. A key
ingredient for these investigations are resonances near the scattering
threshold. Although such resonances have been theoretically
\cite{Bohn:PRA87:2013} and experimentally
\cite{h2co-prl-2012,shd-prl-2012,narevicius-2012} explored in the
chemistry of low temperature systems, their effect on reaction rates
is however not fully taken into account.

In this article, we investigate how reaction and inelastic processes
are affected by near threshold resonances (NTR) in the entrance
channel of a reactive scattering system. We focus our attention on two
benchmark systems containing hydrogen, namely H$_2$+Cl and H$_2$+F.
We note that alkali hydrides \cite{alkali-hydrides}, because of their
small mass, would also be interesting systems to investigate.  By
varying the mass of H, explore the modification of the behavior of the
reaction cross section $\sigma^{\rm react}$ from the Wigner's
threshold law $\sigma^{\rm react}\propto \varepsilon^{-1/2}$
\cite{wigner,hossein} at ultralow energy $\varepsilon$. We generalize
our initial treatment \cite{our-NTR} of the effect of resonances in
low energy collisions originally analyzed in
nuclear~\cite{Bethe:PR:1935} and atomic~\cite{fano} collisions, to
include a multi-channel formalism using Jost functions.

In Section~\ref{sec:general}, we review the general theory of reactive
scattering and discuss the two benchmark systems H$_2$+Cl and H$_2$+F,
and present the results of our calculations in
Section~\ref{sec:results} for both systems. In Section~\ref{sec:Smat},
we give an explanation of those results based on the S-matrix,
followed by an alternative description based on Jost functions in
Section~\ref{sec:Jost}. Finally, we describe the equivalence of both
approaches in Section~\ref{sec:equiv} and give a simple physical
picture in Section~\ref{sec:wave}, before concluding in
Section~\ref{sec:conclusion}.

\section{Reactive scattering at low energies
  \label{sec:general}}

In this section, we review briefly the theory of low energy reactive 
scattering, and describe the two benchmark systems that we use to
study the effect of $s$-wave near threshold resonances.

\subsection{General theory}

The scattering cross
section from an initial internal state $i$ to a final state $f$ is
given by \cite{alex-bala}
\begin{equation}
   \sigma_{f\leftarrow i}(\varepsilon_i) = \frac{\pi}{k_i^2}
   \sum_{J=0}^{\infty} \left( \frac{2J+1}{2j+1}\right) \sum_\ell
   \sum_{\ell'} \left| \delta_{fi} - S_{fi}^J\right|^2,
   \label{eq:full-xsection}
\end{equation}
where $\ell = |J-j|, \dots, J+j$ and $\ell'=|J-j'|,\dots,J+j'$; $\bm J
= \bm j + \bm\ell = \bm j' + \bm\ell'$ is the total angular momentum,
with molecular rotational momentum $\bm j$ and orbital angular $\bm
\ell$ in the entrance channel $i$, and corresponding quantum numbers
$J$, $j$, and $\ell$ (the primes indicate the exit channel $f$). Here,
$\varepsilon_i = \hbar^2 k_i^2/2\mu$ is the kinetic energy with
respect to the entrance channel threshold, $k_i$ the wave number, and
$\mu^{-1}=m_{\mathrm{H}_2}^{-1}+m_{\mathrm{A}}^{-1}$ the reduced mass
in the entrance arrangement (with A standing for Cl or F).  We are
focusing on the effect of resonances at ultralow temperatures, and so
consider only $s$-wave scattering with $\ell =0$, which requires $J=j$
and thus $(2J+1)/(2j+1)=1$. In addition, we limit ourselves to
molecules initially in their rotational ground state ($j=0$), such
that we only need consider $J=0$; consequently, the sum $\sum_{\ell'}$
in (\ref{eq:full-xsection}) reduces to a single term: $\ell'=j'$.
Hence, equation~(\ref{eq:full-xsection}) simplifies significantly, and
it now reads:
\begin{equation} \sigma_{f\leftarrow i}(\varepsilon_i) = \frac{\pi}{k_i^2}
    \left| \delta_{fi} - S_{fi}^{J=0}(k_i)\right|^2. 
    \label{eq:sigma-s-wave}
\end{equation}

It is well known \cite{bala-cplett} that in the zero-energy limit the
cross sections can be expressed in terms of the complex scattering
length $a_i =\alpha_i -i\beta_i$,
\begin{equation}
\begin{array}{lll}
   \sigma_i^{\rm react} (\varepsilon ) \equiv  \displaystyle \sum_{f\neq i} 
   \sigma_{f\leftarrow i}(\varepsilon)
   & \rightarrow & \displaystyle 4\pi\;\frac{\beta_i}{k},\\ 
      \sigma_i^{\rm elast} (\varepsilon )  \equiv  \sigma_{i\leftarrow i}(\varepsilon) 
   & \rightarrow & 4\pi (\alpha_i^2 + \beta_i^2). 
\end{array}
 \label{eq:alpha-beta}
\end{equation}
We obtain energy dependent rate coefficients by multiplying the cross
sections with the relative velocity $v_{\rm rel} = \hbar k/\mu$,
\begin{equation}
\begin{array}{lll}
   K_i^{\rm react} (\varepsilon ) 
       \equiv\displaystyle  \frac{\hbar k}{\mu} \sum_{f\neq i} 
   \sigma_{f\leftarrow i}(\varepsilon)
   & \rightarrow & \displaystyle \frac{4\pi\hbar}{\mu} \beta_i,\\ 
      K_i^{\rm elast} (\varepsilon )  \equiv \displaystyle \frac{\hbar k}{\mu}  
      \sigma_{i\leftarrow i}(\varepsilon) 
   & \rightarrow &\displaystyle \frac{4\pi\hbar k}{\mu} (\alpha_i^2 + \beta_i^2),
\end{array}
 \label{eq:K-alpha-beta}
\end{equation}
which can be thermally averaged over a Maxwellian velocity
distribution to yield the true rate constant $K(T)$.  Note that the
sum $\sum_{f\ne i}$ in Eqs.~(\ref{eq:alpha-beta}) and
(\ref{eq:K-alpha-beta}) includes all possible final states, i.e., both
reaction channels to form the products and quenching (inelastic)
channels for the reactant.  Thus, $\sigma^{\rm react}$ and $K^{\rm
  react}$ defined above include all non-elastic outcomes of the
scattering process, i.e., reaction plus quenching.   We shall
simplify our notation by omitting the channel subscript $i$ (for
$\varepsilon_i$, $k_i$, $\alpha_i$, $\beta_i$, etc.) and we also omit
the superscript $J$ in $S_{fi}^{J}$ throughout the remainder of this
article.

\begin{figure}[hb]
\centerline{
\includegraphics[clip,width=0.9\linewidth]{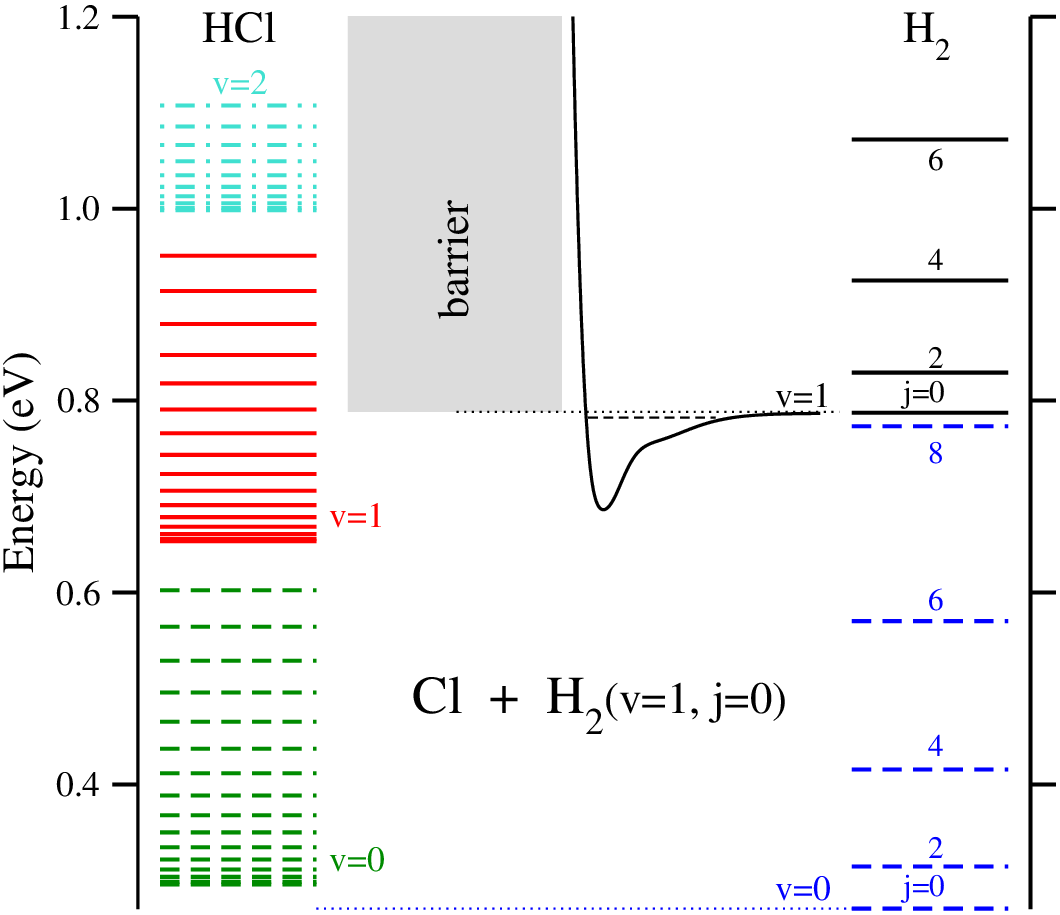}
}
\caption{Rovibrational energy levels for H$_2$ and HCl.  For H$_2$ we
  show the rotational series for $v=0$ (dashed blue) and $v=1$ (solid
  black).  Similarly for HCl, we have $v=0$ (dashed green), $v=1$
  (solid red) and $v=2$ (dot-dashed cyan).  The inset shows a sketch
  of the diagonal matrix element of the potential energy surface for
  the $(v=1,j=0)$ ro-vibrational entrance channel; the dashed line
  represents a weakly quasi-bound level for the van der Waals complex
  Cl$\cdots$H$_2$ in the entrance channel H$_2(1,0)$.}
\label{fig:h2cl-PES}
\end{figure}

We remark that, although the limits in Eq.~(\ref{eq:alpha-beta})
remain valid even in the presence of NTRs, their applicability will be
limited to a much narrower domain of energies, and a new behavior will
emerge for the remainder of the low energy regime, as we will show in
this article.

\subsection{Systems with reaction barrier}

For our study, we selected two barrier-dominated reactive systems,
namely H$_2$ + Cl $\rightarrow$ H + HCl and H$_2$ + F $\rightarrow$ H
+ HF.  Although both have reaction barriers, they have also different
level structure energetics.

As shown in Fig.~\ref{fig:h2cl-PES}, low energy scattering is purely
elastic for Cl + H$_2$ with H$_2$ in its rovibrational ground state.
We therefore consider the next vibrational level as an initial state,
namely $(v=1,j=0)$, such that the reaction to form HCl become possible
even in the ultracold regime.  Note that quenching of H$_2$ by Cl will
also be a possible (inelastic) outcome.  Fig.~\ref{fig:h2cl-PES} also
shows the shallow potential well of the Cl$\cdots$H$_2$ van~der Waals
complex which supports quasibound states leading to resonances in the
entrance arrangement \cite{ClH2:Science:2008}. This system was
recently investigated at ultralow temperatures by Balakrishnan
\cite{Bala:H2Cl:2012} and by us \cite{our-NTR}.  In this work, we use
the potential energy surface developed by Bian and
Werner~\cite{BWpes:JCP:2000}.

\begin{figure}[b]
\centerline{
\includegraphics[clip,width=.9\linewidth]{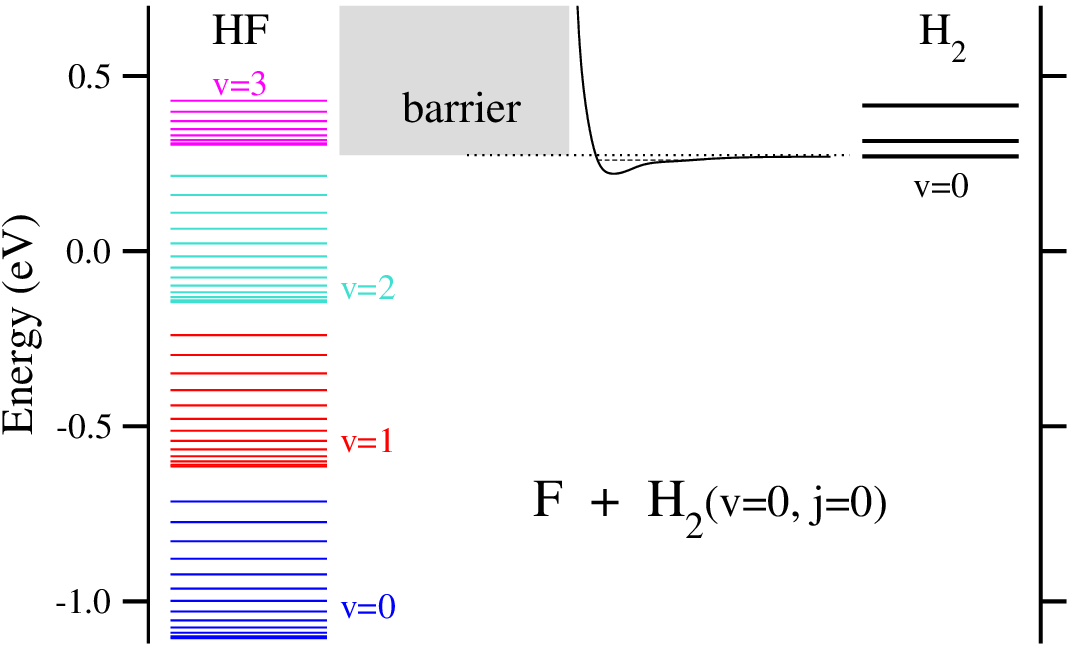}
}
\caption{\label{fig:levels-fh2} Same as Fig.~\ref{fig:h2cl-PES} for
HF. Only $v=0$, $j=0,\ 2,\ 4$ are shown for H$_2$.}
\end{figure}

In the case of H$_2$ + F, reaction channels are already open, and the
formation of HF at ultralow energies is possible even if H$_2$ is in
its rovibrational ground state $(v=0,j=0)$. Note that quenching is no
longer possible in this case.  However, there is a wide range of
rovibrational states available for the product HF, as illustrated in
Fig.~\ref{fig:levels-fh2}.  Again, the shallow potential well in the
entrance channel can support quasibound states of the F$\cdots$H$_2$
van~der Waals complex, and if such a state is near the threshold, it
will produce resonance effects for low energy scattering.  This system
has been studied in the ultracold regime by Dalgarno and co-workers
\cite{H2F-Bala,Bodo:JPB:2004}.  We use the potential energy surface
developed by Stark and Werner~\cite{SWpes}.

As we did in our previous study of H$_2$+Cl \cite{our-NTR}, we
explore the effect of near threshold resonances (NTR) on reactive
scattering by varying the mass of H. This was also done
in~\cite{Bodo:JPB:2004}; the channel thresholds in both arrangements
H$_2$--Cl and H--HCl (or H$_2$--F and H--HF) shift at different rate,
an approach similar to modifying the potential surface
itself~\cite{JMH:JCP:2007}.

\section{Results and discussion}
\label{sec:results}

The results we present here were obtained using the \textsc{abc}
reactive scattering code of Manolopoulos and
coworkers~\cite{ABC:CPC:2000}, which we have optimized for ultralow
energies in previous studies of H$_2$+D \cite{PCCP-H2+D,NJPH2+D,PRLH2+D},
and H$_2$+Cl \cite{our-NTR}.

\begin{figure}[ht]
\centerline{
\includegraphics[clip,width=0.99\linewidth]{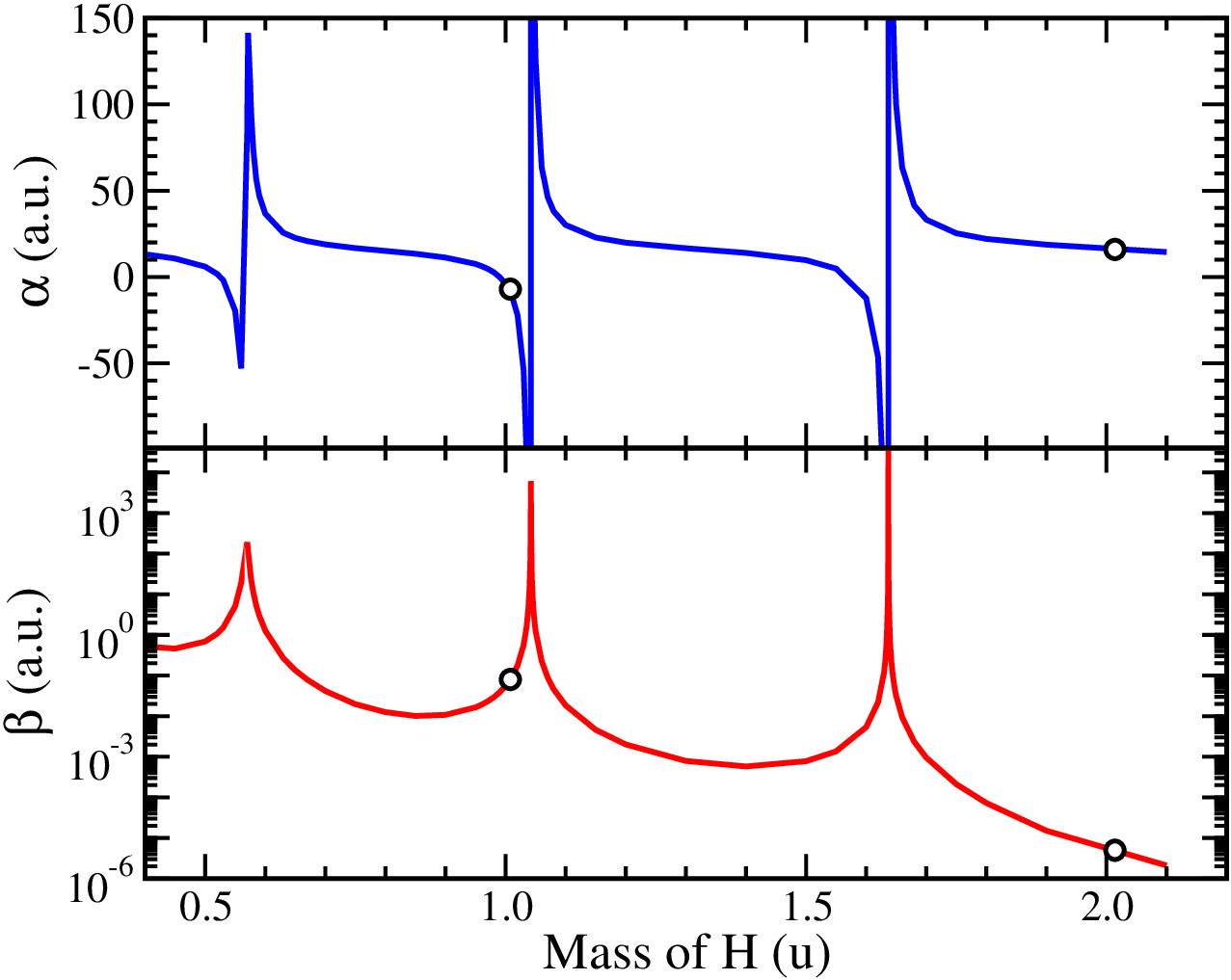}
}
\caption{Real ($\alpha$) and imaginary ($\beta$) components of the
  scattering length as functions of the mass of the hydrogen atom $m$
  for H$_2$+Cl.  The true masses of hydrogen and deuterium are
  indicated by circles.
}
\label{fig:alpha-beta}
\end{figure}

\subsection{Results for H$_2$ + Cl $\rightarrow$ HCl + H}

Figure \ref{fig:alpha-beta} shows the real and imaginary parts of the
scattering length for H$_2(v=1,j=0)$+Cl as a function of the mass $m$
of H, with the open circles indicating the true masses of H and D. We
note that H is located on the wing of a resonance, visible as a sharp
increase for $\beta$.

\begin{figure}[ht]
\includegraphics[clip,width=\linewidth]{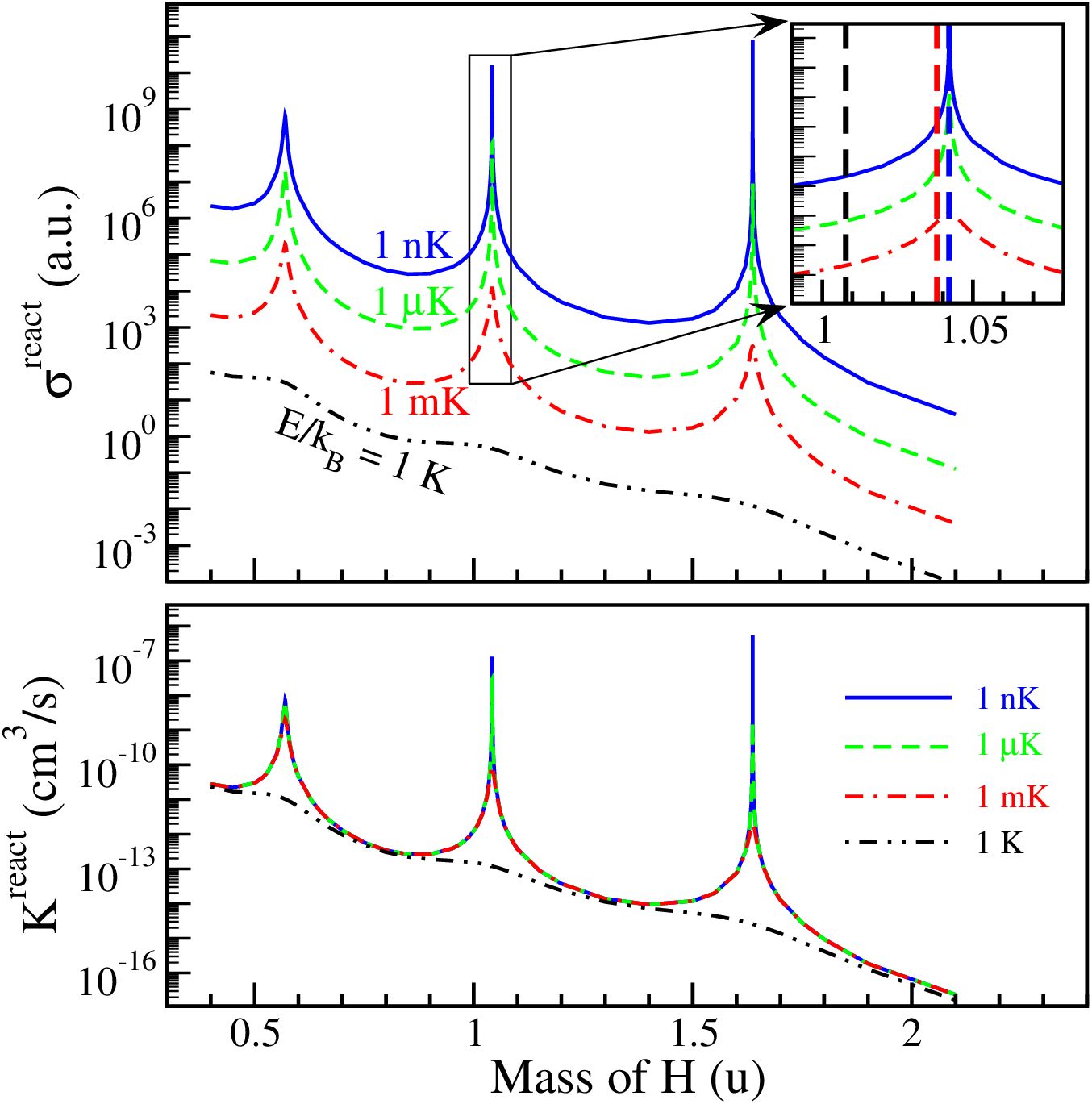}
\caption{\label{fig:clh2-rate-mass} Top panel:
  Reaction cross section for H$_2$+Cl as a function
  of mass $m$ of H for different scattering energies in units of
  Kelvin. The inset shows $\sigma^{\rm react}$ in the vicinity of the
  resonance with the vertical dashed lines corresponding to
  $m=1.0078\,\mathrm{u}=m_{\rm{H}}$ (true mass, in black), 1.038\,u
  (red), and 1.042\,u (blue), from left to right respectively.  Bottom
  panel: Rate constant for the same energies.}
\end{figure}

The top panel of Fig.~\ref{fig:clh2-rate-mass} shows the reaction
cross section as a function of $m$ for a few energies (in Kelvin).
The simple $\sigma\propto k^{-1} \propto \varepsilon^{-1/2}$ scaling
in Eq.~(\ref{eq:alpha-beta}) implies equidistant curves for the
energies chosen in the logarithmic scale.  We find this to be true
except near the resonances. This is more clearly illustrated in the
lower panel of Fig.~\ref{fig:clh2-rate-mass} in which the rate
constant is plotted; from Eq.(\ref{eq:K-alpha-beta}), we expect curves
to coincide for all those four low energies, which is the case away
from the resonances. However, as we near a resonance, the difference
between the curves become more pronounced.

\begin{figure}[h]
\centerline{\includegraphics[clip,width=.7\linewidth]{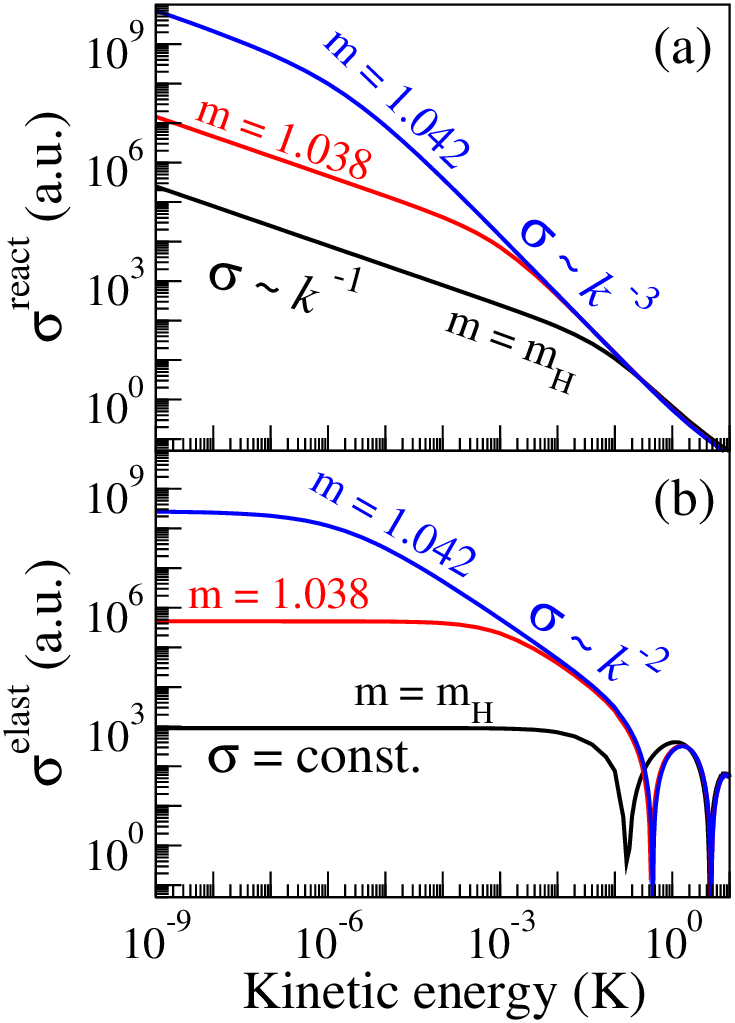}}
\caption{\label{fig:ClH2xsec} Energy dependence of the reaction cross 
  section (a) and elastic cross section (b).   The results 
  shown are for the entrance channel  H$_2(v\!\!=\!\!1,\,j\!\!=\!0)$ + Cl
  $\rightarrow$ H + HCl, and they correspond to different masses of
  H, as indicated for each curve.}
\end{figure}

In order to analyze this behavior, we selected three masses near the
resonance, namely $m=1.0078$\,u$=m_{\rm H}$ (true mass), 1.038\,u, and
1.042\,u; they appear as dashed vertical lines in the inset of the top
panel of Fig.~\ref{fig:clh2-rate-mass}. We focus our attention on
these three masses, and we analyze in detail the energy dependence of
the cross sections. Fig.~\ref{fig:ClH2xsec}(a) shows the low energy
behavior of $\sigma^{\rm react}$: as $\varepsilon\rightarrow 0$,
$\sigma^{\rm react}$ reaches the Wigner regime, scaling as
$\varepsilon^{-1/2}$ for all three masses, but for masses closer to
the resonance, the scaling changes to $\varepsilon^{-3/2}$.
Fig.~\ref{fig:ClH2xsec}(b) shows the elastic cross sections
$\sigma^{\rm elast}$ for the same masses; the Wigner regime's constant
cross section as $\varepsilon\rightarrow 0$ changes to the expected
$\varepsilon^{-1}$ scaling for $m$ near a resonance.

\begin{figure}[ht]
\centerline{\includegraphics[clip,width=.7\linewidth]{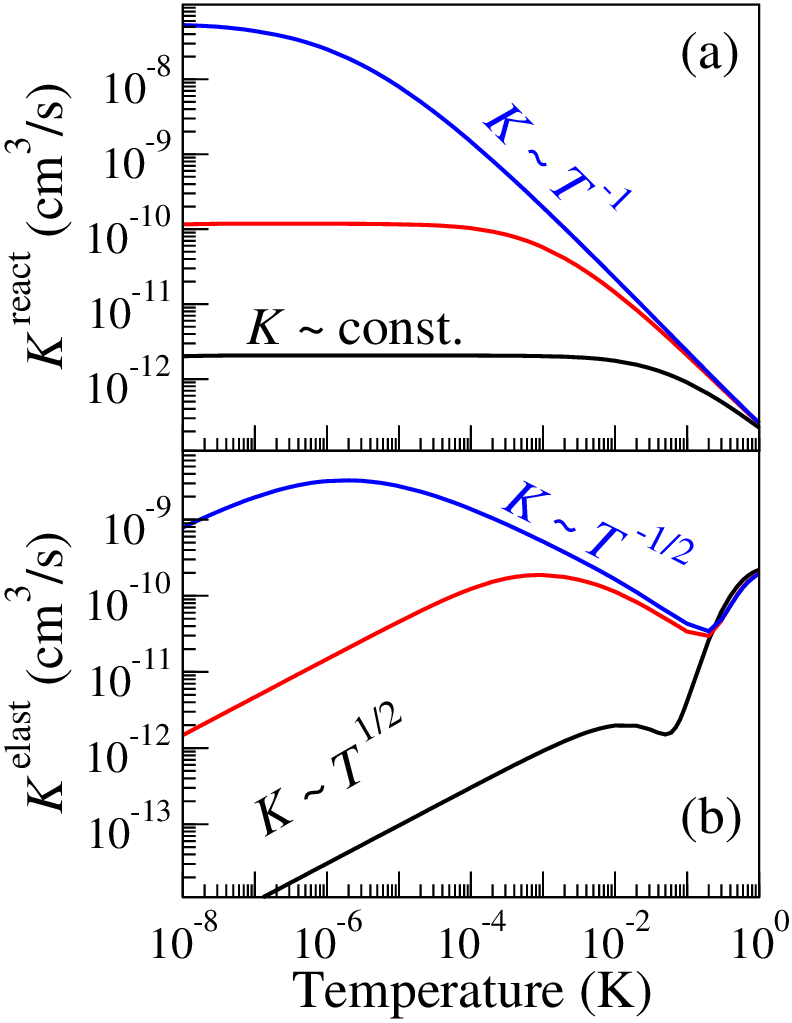}}
\caption{\label{fig:ClH2rate}   Temperature dependence of 
  rate coefficients for reaction (a) and elastic scattering (b)
  corresponding to the cross sections shown in Fig.~\ref{fig:ClH2xsec}.}
\end{figure}

For the sake of completeness, the corresponding rate constants are
plotted in Figs.~\ref{fig:ClH2rate}(a) and (b). As $m$ nears the
resonance, $K^{\rm react}$ is significantly enhanced, scaling as
$T^{-1}$ until $T$ is small enough that the Wigner regime is reached
and $K^{\rm react}$ becomes constant. The corresponding scaling for
$K^{\rm elast}$ changes from $T^{-1/2}$ for NTR to $T^{1/2}$ for
Wigner's regime.

\subsection{ Results for H$_2$ + F $\rightarrow$ HF + H}

\begin{figure}[t]
\centerline{
\includegraphics[clip,width=.95\linewidth]{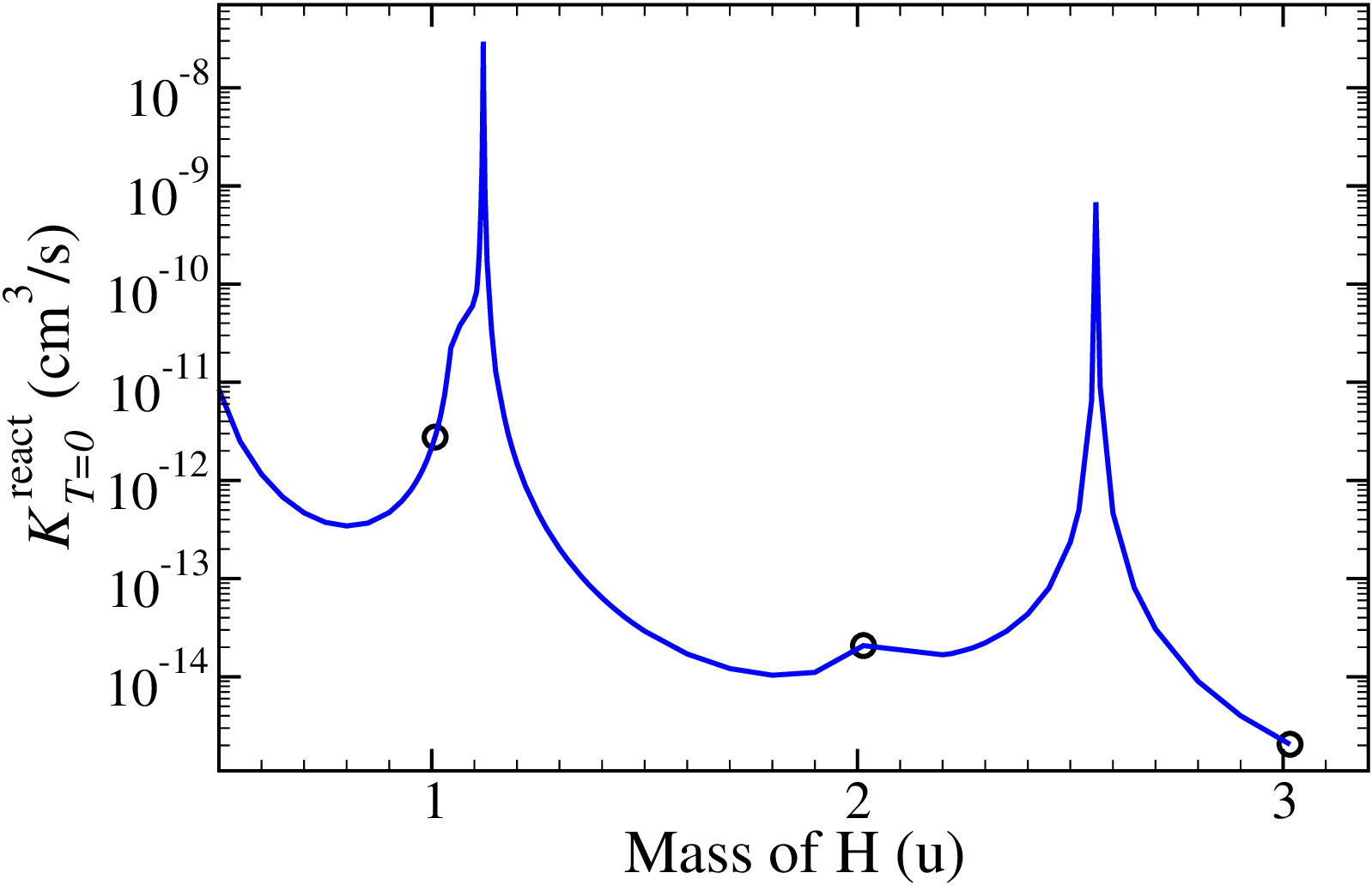}
}
\caption{\label{fig:rate-mass-h2f}Mass dependence of the reaction rate
  coefficient for F + H$_2$(0,0) $\rightarrow$ HF + H at $T=10^{-6}$K.}
\end{figure}

The results for the F + H$_2(0,0)$ reaction contain resonant features
that are similar to those described above for the Cl + H$_2(1,0)$
system.  Figure~\ref{fig:rate-mass-h2f} shows the reaction rate
constant at $T=0$, which is directly related to $\beta$, see
Eq.~(\ref{eq:K-alpha-beta}).  Although Fig.~\ref{fig:rate-mass-h2f}
and Fig.~\ref{fig:alpha-beta} are similar, only two pronounced
resonances are found for H$_2$ + F, even though the mass of H varies
over a wider range.  Note that the real mass of H is again on the wing
of a resonance.  Dalgarno and co-workers \cite{Bodo:JPB:2004} have
obtained similar results, but their work has explored a limited mass
range (only up to $m=1.5$\,u, with one additional data point for
$m=m_D\approx 2$\,u) and they only found one resonance (near
$m=1.1$\,u).

\begin{figure}[t]
\centerline{
\includegraphics[clip,width=.95\linewidth]{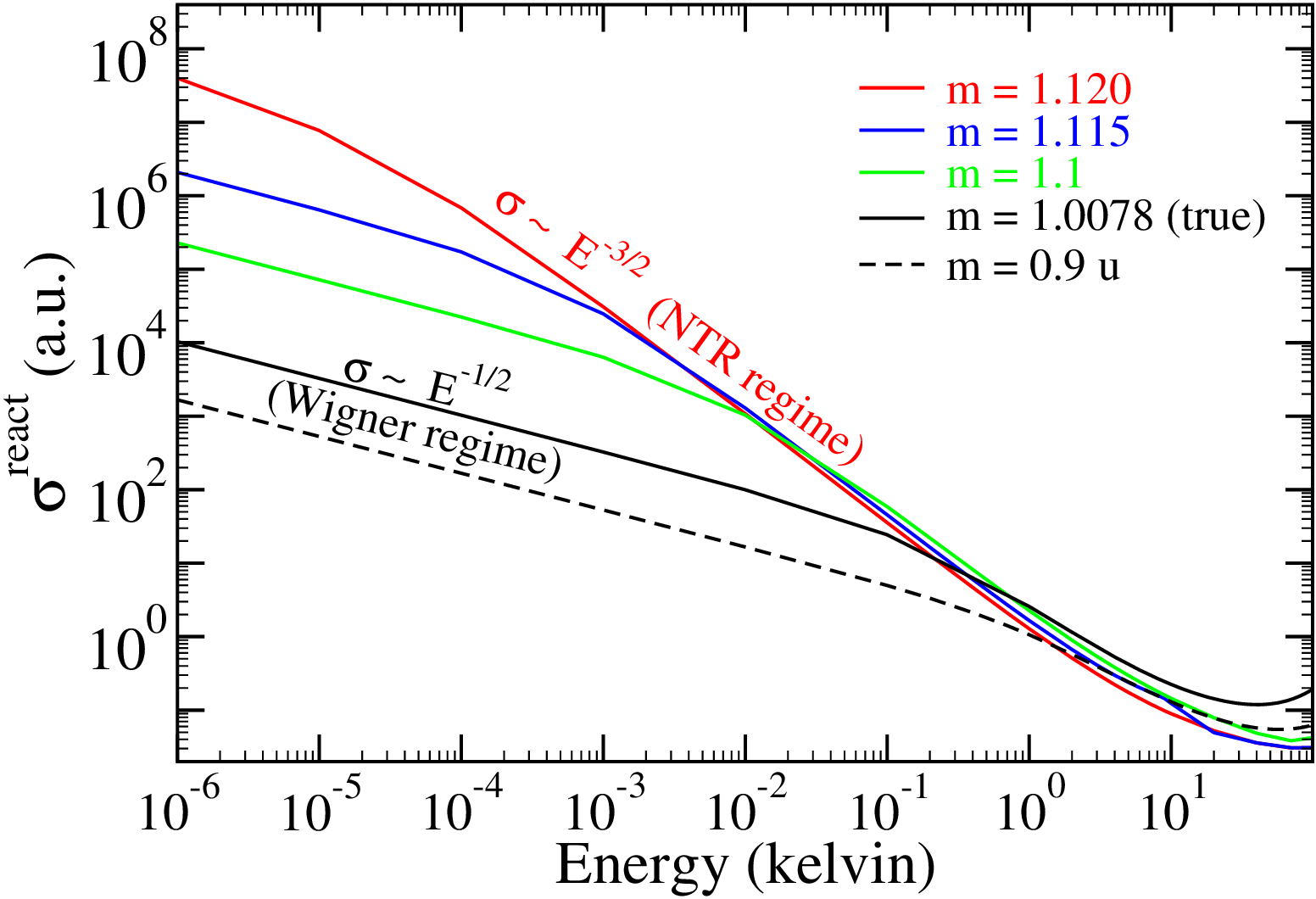}
}
\caption{\label{fig:sigma-E-H2F} Energy dependence of the reaction
cross section for F + H$_2(0,0)$ for different masses of H.}
\end{figure}

\begin{figure}[b]
\centerline{
\includegraphics[clip,width=.95\linewidth]{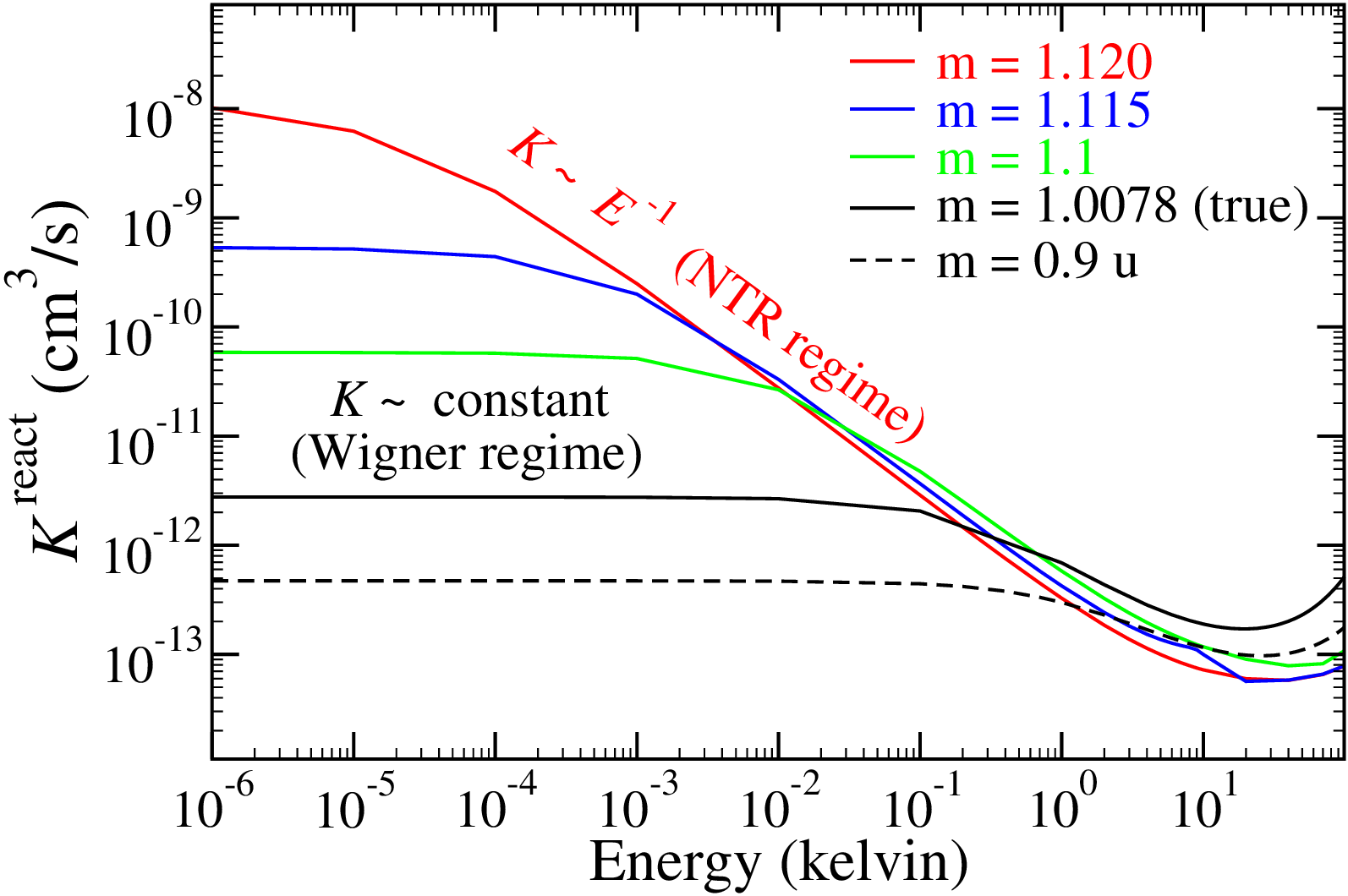}
}
\caption{\label{fig:rate-E-H2F} Energy dependence of the reaction rate
constant for F + H$_2(0,0)$ for different masses of H.}
\end{figure}

The reaction cross section and rate constant follow the same behavior
as the mass gets closer to the resonance. This is illustrated in
Fig.~\ref{fig:sigma-E-H2F} for $\sigma^{\rm react}$, with the NTR
scaling ($\varepsilon^{-3/2}$) becoming more apparent as $m$ gets
closer to the resonance, while the Wigner regime (with
$\varepsilon^{-1/2}$ scaling) is pushed to lower energies. The
corresponding behavior for the rate constant is shown in
Fig.~\ref{fig:rate-E-H2F}.

To explain the appearance of the NTR regime, we turn our attention to 
the analytical properties of the $S$-matrix.

\section{ S-matrix approach for $\ell=0$}
\label{sec:Smat}

The anomalous behavior of the cross sections shown above---an abrupt
increase followed by a gradual transition into the Wigner regime, as
the energy decreases---is due to the presence of a resonance pole near
the threshold of the entrance channel. To understand this, we pay
attention to the $S$~matrix.  Recalling the unitarity condition,
$\sum_f |S_{fi}|^2 = |S_{ii}|^2 + \sum_{f\neq i} |S_{fi}|^2 =1$, and
using Eq.~(\ref{eq:sigma-s-wave}), we rewrite the cross sections in
Eq.~(\ref{eq:alpha-beta}) explicitly in terms of the diagonal element
$S_{ii}(k)$:
\begin{eqnarray}
 & & \sigma^{\rm react}(k) =  \frac{\pi}{k^2} \left(1-|S_{ii}(k)|^2\right),
                                         \label{eq:sigma-in-S-matrix} \\
 & & \sigma^{\rm elast}(k) = \frac{\pi}{k^2} |1-S_{ii}(k)|^2.
                                         \label{eq:sigma-el-S-matrix}
\end{eqnarray}

In this section we use the single channel case heuristically to guide
our analysis of $S_{ii}(k)$, while in the next section we will present
a more rigorous approach based on Jost matrices for the general case
of a many channel problem.

\subsection{Pole factorization; single channel case}

For a single channel problem, we express the S matrix for a given
partial wave $\ell$, in terms of the Jost function $\mathcal{F}\!\!_\ell$:
\begin{equation}
   S_\ell(k) = \frac{\mathcal F\!\!_\ell(-k)} {\mathcal F\!\!_\ell(k)}.
\label{eq:Sjost}
\end{equation}
Thus, the properties of $S_\ell(k)$ follow from those of the Jost
function \cite{taylor}; here, we shall make use of the fact that if
$\mathcal{F}\!\!_\ell(k)$  has a zero located at $k=p$ (in the
complex plane of the momentum), then the $S$-matrix has a pole at
$k=p$ and a zero at $k=-p$.  Note that for any partial wave $\ell\geq
1$, two poles will approach the threshold and collide at $k=0$, while
for s-wave ($\ell=0$) only a single pole at a time may reach the
threshold.  In this paper we analyze the case of s-wave scattering,
for which it is convenient to factor out the effect of the pole at
$k=p$ and its accompanying zero at $k=-p$.  Thus, we can write
\begin{equation}
\label{eq:Sfact}
   S(k) = \frac{p+k}{p-k}\tilde{S}(k),
\end{equation}
and we assume the background contribution $\tilde S(k)$ is a slowly
varying function of $k$.  Note that we shall omit the subscript
$\ell$, as we only discuss the case of $\ell=0$.

As is well known \cite{taylor}, for purely elastic scattering in
partial wave $\ell=0$, a pole $p$ which is sufficiently close to $k=0$
will be located on the imaginary axis; thus, we can write
\begin{equation}\label{eq:p=ikappa}
p = i\kappa ,
\end{equation}
with $\kappa$ real valued; when $\kappa>0$ the pole corresponds to a
bound state, while for $\kappa<0$ it corresponds to a virtual (or
anti-bound) state.  Since we are interested in near threshold
resonances, we shall focus our attention on a short segment of the
trajectory of the pole along the imaginary axis near $k=0$.  From
Eqs.~(\ref{eq:p=ikappa}) and (\ref{eq:Sfact}) it is clear that the
resonant contribution $S^{\mathrm{res}}(k) \equiv \frac{p+k}{p-k}
=\frac{i\kappa+k}{i\kappa-k}$ is explicitly unitary for real values
of $k$.  The full $S$ matrix is also unitary for real $k$, and thus
the background part has to be unitary, and both of them can be written
in the usual fashion, i.e., $S=e^{2i\delta}$ and
$\tilde{S}=e^{2i\tilde\delta}$.  The full phaseshift $\delta$ is a sum
of the background phaseshift $\tilde{\delta}$ and the resonant
contribution,
\[
\delta(k)=\tilde\delta(k)-\arctan\left({\textstyle\frac{k}{\kappa}}\right).
\]
Hence, the elastic cross section,
$\sigma^{\mathrm{elast}}=\frac{4\pi}{k^2}\sin^2\delta$, reads
\[
\sigma^{\rm elast}(k) = \frac{4\pi}{k^2}\,
   \frac{(k\cos\tilde\delta - \kappa\sin\tilde\delta)^2}
        {k^2 + \kappa^2}.
\]
At low energy we use the approximation
$\sin\tilde\delta\approx\tilde\delta(k)\approx-k\tilde a$ (and
$\cos\tilde\delta\approx 1$) to further simplify the expression above:
\begin{equation}\label{eq:sigma-el-kappa}
\sigma^{\rm elast}(k) \approx 4\pi
   \frac{ (1 + \kappa\tilde a)^2 }
        {k^2 + \kappa^2}=
   4\pi a^2 \frac{ \kappa^2 }
                 {k^2 + \kappa^2} ,
\end{equation}
where $\tilde a$ is the background scattering length, and
$a=\tilde{a}+\frac{1}{\kappa}$ the full scattering length.  We remark
that the Lorentzian expression in the equation above is a function of
the \emph{momentum} $k$.  Thus, although
Eq.~(\ref{eq:sigma-el-kappa}) is similar to the Breit--Wigner
formula, we recall that the latter has the familiar Lorentz-type
\emph{energy} dependence.  Moreover, since the lineshape in
Eq.~(\ref{eq:sigma-el-kappa}) is centered on the threshold ($k=0$)
itself, only the $k>0$ half can be probed in elastic scattering, which
is specific to the case of a near threshold resonance for partial wave
$\ell=0$ in the entrance channel.

When the resonant pole crosses the threshold, the scattering length
diverges:
\begin{equation}\label{eq:a-kappa}
a = \tilde a + \kappa^{-1} 
   \  \xrightarrow{\ \kappa\rightarrow\,\pm 0\ } \ \pm\infty.
\end{equation}
Hence, the behavior of the cross section will be affected
dramatically; indeed, when $\kappa$ is vanishingly small,
the denominator in Eq.~(\ref{eq:sigma-el-kappa}) will give
rise to two different types of behavior inside the low energy domain,
depending on the relative magnitude of $\kappa$ and $k$.  When
$k\ll|\kappa|$ we recover the familiar result for the Wigner regime,
\[
\sigma^{\rm elast}(k) \ \xrightarrow{\ k\rightarrow\,0\ } \  \  4\pi a^2 ,
\]
while for the remainder of the low energy domain, $\kappa$ becomes
negligible when $k\gg|\kappa|$, and we obtain
\[
\sigma^{\rm elast}(k) \ \xrightarrow[\ k\,\gg\,|\kappa|\ ]{} \  4\pi
     \frac{(1 + \kappa\tilde a)^2}{k^2}
\ \ \xrightarrow[\ \kappa\,\approx\,0\ ]{}\ \  \frac{4\pi}{k^2},
\]
which we refer to as \emph{NTR-regime behavior}.  From our discussion,
it is clear that the extent of the NTR regime is specified
approximately by the inequalities $|\kappa|\lessapprox
k\lessapprox|\tilde{a}|^{-1}$.

We emphasize that although the Wigner regime can be displaced
towards the extreme low energy domain, Wigner's threshold law is
always recovered when $k\rightarrow 0$, except in the critical case of
$\kappa=0$, which corresponds to a bona fide zero-energy resonance.
Only in this special case, could the Wigner behavior be lost, and
$\sigma^{\mathrm{elast}}(k)$ would diverge at $k=0$, since it would
follow the NTR-regime behavior throughout the entire low-$k$ regime.
As we shall see in the next section, the special case of a zero-energy
resonance is only possible for purely elastic scattering, because any
additional (inelastic or reactive) open channels---which are coupled,
however feebly, to the entrance channel---will push the resonance pole
away from the imaginary axis; hence, the pole will be kept away from $k=0$,
but still possibly close enough to produce a significant resonance
effect.

\subsection{Pole factorization; many channel case}

We consider first the reaction cross section, followed by the
elastic cross section.

\subsubsection{The reaction cross section}

We now extend the analysis of the single channel $S$ matrix to the
many channel case; namely, we focus on the diagonal element
$S_{ii}(k)$ corresponding to the entrance channel.  It is important to
note that the factorization employed in the single channel case in
Eq.~(\ref{eq:Sfact}) also holds for the diagonal element corresponding
to the entrance channel; indeed, the general properties of the $S$
matrix ensure that a pole at $k=p$ is always accompanied by a zero of
$S_{ii}(k)$ at $k=-p$ \cite{newton:0}.  Thus, for a NTR in the
entrance channel $i$, we write
\begin{equation}
   S_{ii}(k) = \frac{p+k}{p-k}\tilde{S}_{ii}(k) .
   \label{eq:Sii}
\end{equation}
The background contribution $\tilde{S}_{ii}(k)=e^{2i\tilde\delta_i}$
is again assumed to be a slowly varying function of $k$.  However,
unlike the single-channel case, the background phase shift
$\tilde{\delta}_i$ is now a complex quantity (owing to the nonzero
reactivity of the system).  Moreover, the pole $p$ is no longer on the
imaginary axis, as will be illustrated by our results, and can be
explained by the fact that a quasi-bound state of the van~der Waals
complex near the threshold of the entrance channel can decay in all
available (open) channels, and thus has a finite lifetime and a finite
decay rate. Should the pole be located on the imaginary axis in the
momentum plane, it would then be located on the real energy axis in
the energy plane, implying a zero width (decay rate $\Gamma=0$) and
thus an infinite lifetime of complex, which would
contradict the assumption of nonzero reactivity.

Our results will show that Re$(p)$ is always negative; thus, we write
explicitly:
\[
p=-q+i\kappa,
\]
with $q>0$.  The nonzero value of $q$ stems from the nonzero reactivity
(inelasticity) of the system, as argued above.  Hence, the
eigen-energy $E_p=p^2/2\mu$ associated with the pole is now a complex
quantity, with Re$(E_p)=(q^2-\kappa^2)/2\mu$ and
Im$(E_p)=-q\kappa/\mu$.  The algebraic sign of $\kappa$ will determine
if the pole represents a virtual state or a quasi-bound state of the
van der Waals complex; indeed, just as in the single channel case,
$\kappa>0$ corresponds to the latter, while $\kappa<0$ to the former.
We remark that the decay rate
$\Gamma=-2\,\mathrm{Im}(E_p)=2q\kappa/\mu$ is positive for a
quasi-bound van~der~Waals complex, while for a virtual state we have
$\Gamma<0$ (unphysical).

At low energy, we employ the usual parametrization of the background
contribution, $\tilde{\delta}_i(k) \approx -k \tilde{a}_i$, where the
background scattering length
$\tilde{a}_i=\tilde{\alpha}_i-i\tilde{\beta}_i$ is now a complex
quantity.  Using similar parametrizations, the resonant part and the
full S matrix element in Eq.~(\ref{eq:Sii}) can be linearized when
$k\to0$.  Thus we have
$S_{ii}^{\mathrm{res}}(k)=\frac{p+k}{p-k}\approx
1-2ia_i^{\mathrm{res}}k$, and
$S_{ii}(k)=S_{ii}^{\mathrm{res}}(k)\tilde S_{ii}(k)\approx 1-2ia_ik$,
and we can extract the resonant contribution to the scattering length,
$a_i^{\mathrm{res}}=\frac i p =\frac{\kappa-iq}{q^2+\kappa^2}$, and
the full scattering length $a_i=\tilde a_i + a_i^{\mathrm{res}}=\tilde
a_i+\frac i p$.  Omitting the channel subscript, the real and
imaginary parts of the scattering length read:
\begin{equation}
\label{eq:alpha}
  \mathrm{Re}(a) \equiv \alpha
    =   \tilde{\alpha} + \frac{\kappa}{q^2+\kappa^2},
\end{equation}
\begin{equation}
\label{eq:beta}
  -\mathrm{Im}(a) \equiv \beta
   =   \tilde{\beta}  + \frac{  q  } {q^2+\kappa^2}.
\end{equation}

We now return to the \emph{unapproximated} expression of the
resonant contribution in Eq.~(\ref{eq:Sii}),
$S^{\mathrm{res}}(k)=\frac{p+k}{p-k}$, and the full S matrix element,
$S(k)=\frac{p+k}{p-k}e^{-2ik\tilde{a}}$, which are valid for the
entire low-energy domain, namely $k\lessapprox|\tilde a|$.
Making use of
$|\tilde{S}(k)|^2\approx|e^{-2ik\tilde a}|^2=e^{-4k\tilde{\beta}}$,
and substituting Eq.~(\ref{eq:Sii}) in
Eq.~(\ref{eq:sigma-in-S-matrix}),
the reaction cross section reads:
\begin{equation}
   \sigma^{\rm react} = \frac{\pi}{k^2} \frac{2e^{-2k\tilde{\beta}}}{|p-k|^2} 
   \left[ (k^2 + |p|^2) \sinh\,2k\tilde{\beta}
   + 2 q k \cosh\,2k\tilde{\beta}) \right] ,
   \label{eq:sigma_in-full}
\end{equation}
Assuming low reactivity, i.e., $\tilde{\beta}\to0$, this result can be
further simplified:
\begin{equation}
\label{eq:sigma-r-beta}
\sigma^{\mathrm{react}}(k) \approx 4\pi \frac{\beta}{k}\,
   \left|\frac{p} {p-k}\right|^2 ,
\end{equation}
with $\beta$ given in Eq.~(\ref{eq:beta}).  Finally, if one neglects
$\tilde{\beta}$ entirely, one can substitute $\beta\approx \beta^{\rm
  res}=q|p|^{-2}$ in the equation above, which can be recast as
\begin{equation}
\label{eq:sigma-r-q}
\sigma^{\mathrm{react}}(k) \approx \frac{4\pi} k \,
   \frac{q} {(q+k)^2 + \kappa^2}.
\end{equation}
This expression only contains two parameters ($q$ and $\kappa$) which
are sufficient to describe a prominent NTR (when $\tilde{\beta}$ is
small).  Note that the full expression given in
Eq.~(\ref{eq:sigma_in-full}) includes the background contribution
$\tilde{\beta}$ as a third parameter, which allows for slight
deviations from the simple $k$-dependence in Eq.~(\ref{eq:sigma-r-q}).

\begin{figure}[t]
\includegraphics[clip,width=\linewidth]{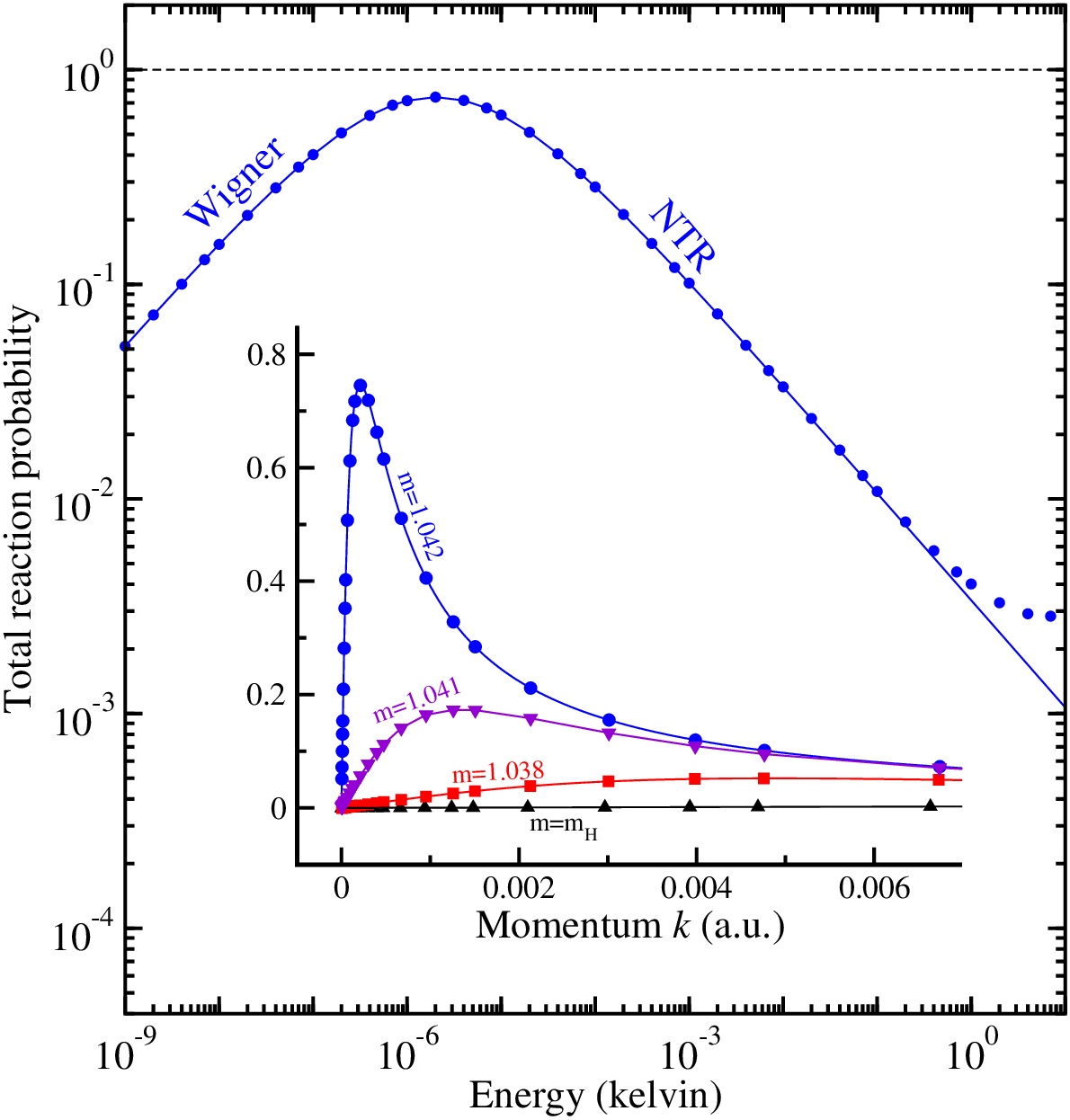}
\caption{\label{fig:smat} Low energy behavior of the total reaction
  probability for $m=1.042$\,u in H$_2$+Cl.  The horizontal dashed
  line marks the unitarity limit.  Inset: linear-scale plot of
  $P^{\mathrm{react}}(k)$ for different masses (as indicated).  The
  data points from the full (exact) computation are shown as symbols,
  while the continuous lines were obtained by fitting using
  Eq.~(\ref{eq:prob}).  }
\end{figure}

Recall that we are interested in the case when $|p|$ is very small,
such that the energy associated with the pole,
$|E_p|=\frac{1}{2\mu}\hbar^2|p|^2$, is well within the ultracold
domain.  Consequently, there will be two different types of behavior
at low $k$, depending on the relative values of $k$ and $|p|$.
Indeed, for $k\ll|p|$, we recover Wigner's threshold law,
\begin{equation}
   \sigma^{\rm react} 
    \ \xrightarrow[\ k\rightarrow\,0\ ]\ \ 4\pi \frac\beta k,
    \label{eq:wigner}
\end{equation}
while for $k\gg|p|$, Eq.~(\ref{eq:sigma-r-q}) reduces to
\begin{equation}
    \sigma^{\rm react} 
    \ \xrightarrow[\ k\,\gg|p|\ ]\ \ 4\pi\frac{q}{k^3},
    \label{eq:NTR} 
\end{equation}
which we named NTR-regime behavior.

We emphasize that the simple expression in Eq.~(\ref{eq:sigma-r-q})
captures both types of behavior, including the transition between the
two regimes, and we can employ it to fit our computed data.
Specifically, we use a more convenient quantity than the cross section
shown in Fig.~\ref{fig:ClH2xsec}(a), namely the reaction probability,
$P^{\mathrm{react}}\equiv 1-|S_{ii}|^2=
\frac{k^2}\pi\sigma^{\mathrm{react}} (k)$, which is bound between zero
and one.  Hence, our fitting procedure employs the expression
\begin{equation}
\label{eq:prob}
P^{\mathrm{react}}(k) = 4q\,
   \frac{k} {(q+k)^2 + \kappa^2},
\end{equation}
which is equivalent with Eq.~(\ref{eq:sigma-r-q}).  The asymmetrical
profile of $P^{\mathrm{react}}(k)$ is clearly illustrated by the
linear-scale plot shown as an inset in Fig.~\ref{fig:smat}, which
contains results for various masses, ranging from the true mass of
hydrogen ($m=m_{\rm H}$) to $m=1.042$\,u.  The results for one of the
strongly resonant cases ($m=1.042$\,u) are also shown on a log-log
scale (see the main plot in Fig.~\ref{fig:smat}), which emphasizes the
good agreement between the numerical results and the simple analytical
expression in Eq.~(\ref{eq:prob}).  Also, the two power-laws in
Eqs.~(\ref{eq:wigner}) and (\ref{eq:NTR}) are easily recognized on the
log-log plot in Fig.~\ref{fig:smat}; indeed, we have
$P^{\mathrm{react}}(k)\sim k$ in the Wigner regime, and
$P^{\mathrm{react}}(k)\sim k^{-1}$ in the NTR regime.

From Eq.~(\ref{eq:prob}) it follows that $P^{\mathrm{react}}(k)$
reaches its maximum value at $k_{\mathrm{max}}=|p|$, which marks the
transition between the two regimes; in terms of energy, the maximum of
$P^{\mathrm{react}}$ is located at
$E_{\mathrm{max}}=|E_p|=\frac{1}{2\mu}(q^2+\kappa^2)$.  Although it is
tempting to interpret the location of the maximum of
$P^{\mathrm{react}}(k)$ as the position of the resonance, one has to
resist this temptation.  Indeed, based on Eq.~(\ref{eq:prob}), we
suggest that it is more appropriate to assign $k=-q$ as the position
of the resonance, instead of $k=k_{\max}=|p|$.

\begin{figure}
\includegraphics[clip,width=\linewidth]{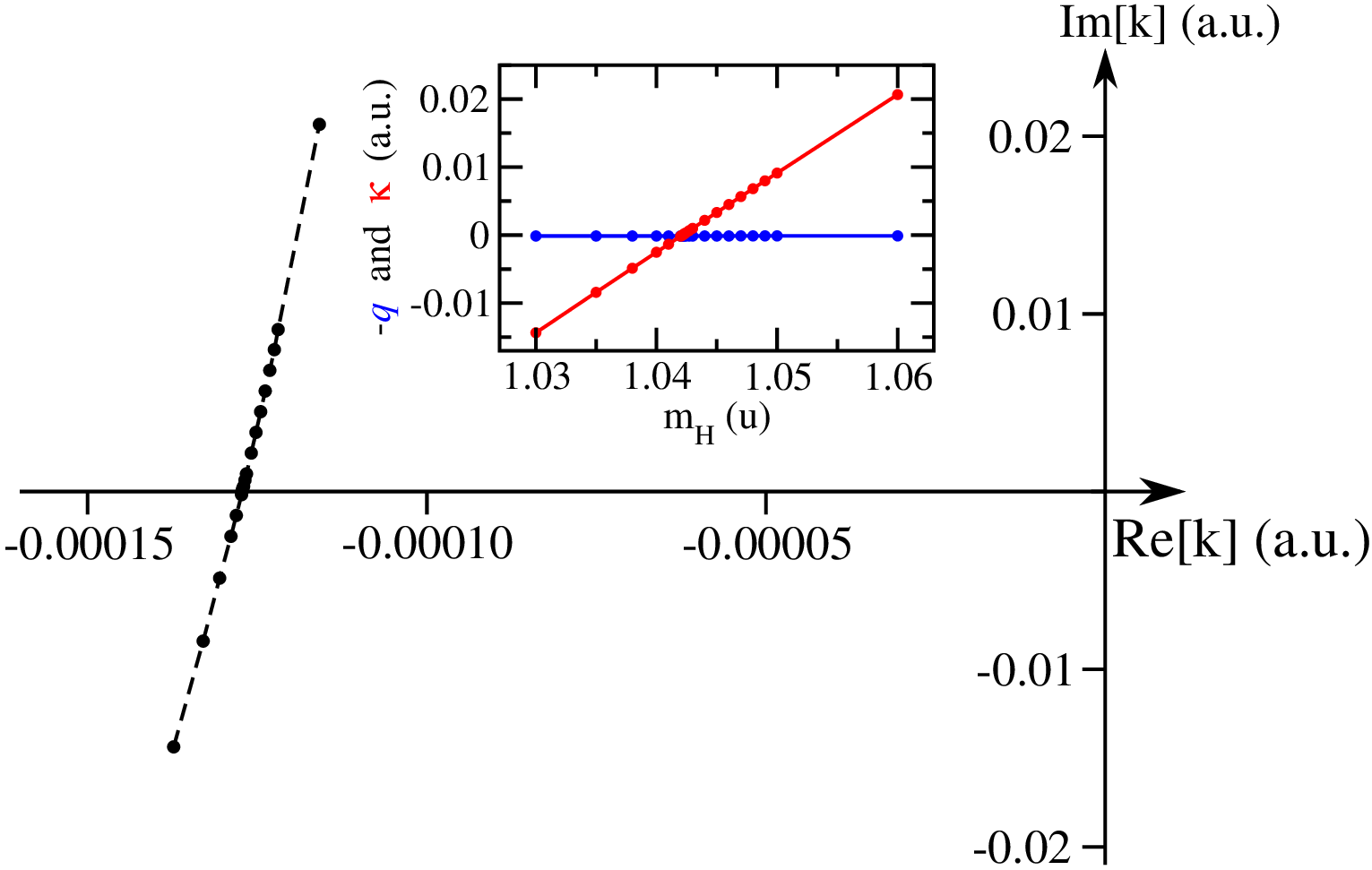}
\caption{\label{fig:pole} Pole trajectory in the complex plane of the
  momentum for H$_2$ + Cl.  The inset shows the explicit mass
  dependence of $q=-\mathrm{Re}[p]$ and $\kappa=\mathrm{Im}[p]$.  Note
  that $q(m)$ is nearly constant, while $\kappa(m)$ has an almost
  linear behavior.  Also note that $|\kappa|$ increases considerably on
  either side of the resonance, such that $q\ll|\kappa|$ (except near
  the very center of the resonance, where $\kappa\approx 0$).}
\end{figure}

We remark that the unitarity limit ($P^{\mathrm{react}}=1$) can only
be reached in the special case with $\kappa=0$, when the pole is
situated on the negative real axis at $p=-q$, and the diagonal element
of $S$ corresponding to the entrance channel has a zero on the
positive real axis at $k=-p=q$.  Namely, we have $S_{ii}(q)=0$, see
Eq.~(\ref{eq:Sii}), which guarantees that $P^{\mathrm{react}}(q)=1$.
Figure~\ref{fig:pole} shows the pole trajectory in the complex plane
as a function of the parameter $m$, and makes it apparent that the
trajectory $p(m)$ crosses the real axis when $m$ reaches a critical
(resonant) value $m=m_{\mathrm{res}}$.  We note that in the special
case of $m=m_{\mathrm{res}}$, when the unitarity limit
$P^{\mathrm{react}}(k)=1$ is reached at $k=q(m_{\mathrm{res}})$, the
NTR regime extends to the lowest possible energies into the ultracold
regime: $k\lessapprox|p(m_{\mathrm{res}})|=q(m_{\mathrm{res}})$.  When
$m$ moves gradually away from $m_{\mathrm{res}}$ the NTR regime
shrinks gradually, and it spans a narrower range of energies (as seen
in Figs.~5, 6, 8, and 9).  Eventually, the NTR regime disappears when
the pole $p(m)$ is no longer in the vicinity of $k=0$.

The pole parameters $q=-\mathrm{Re}(p)$ and $\kappa=\mathrm{Im}(p)$
are treated as fitting variables, and they are thus obtained from the
computed results.  Note that the value of $\tilde{\beta}$ can only
be extracted from the fitting procedure if the full expression in
Eq.~(\ref{eq:sigma_in-full}) is employed instead of the simpler
Eq.~(\ref{eq:sigma-r-q}); we also remark that when $\tilde{\beta}$
is negligibly small compared to the resonant contribution
$\beta^{\mathrm{res}}=q(q^2+\kappa^2)^{-1}$, see Eq.~(\ref{eq:beta}),
then it can no longer be extracted reliably from the fitted data.

We emphasize that we conducted our analysis in the complex plane of
the momentum $k$, rather than the complex energy plane; to appreciate
the unsuitability of the latter, we simply point out that the complex
energies associated with both the pole ($k=p$) and the zero ($k=-p$)
are identical,
\begin{equation} 
   E_p = \frac{\hbar^2 (\pm p)^2}{2\mu}
     \equiv \varepsilon_p - i\hbar\frac{\Gamma}{2} \;,
\end{equation}
with $\varepsilon_p = \frac{\hbar^2}{2\mu}(q^2-\kappa^2)$ and $\Gamma
= \frac{2\hbar}{\mu} q\kappa$.  We recall that one must have $q>0$ for
a physical cross section, see Eq.~(\ref{eq:sigma-r-q}), and thus we
have $\mathrm{sgn}(\Gamma)=\mathrm{sgn}(\kappa)$.  Hence, for a quasi-bound
vdW-complex ($\kappa>0$) we have $\Gamma>0$, while for the case of a
virtual state ($\kappa<0$) we have $\Gamma < 0$ (unphysical).
Finally, we remark that for an $s$-wave NTR, scattering alone cannot
distinguish between shallow bound and anti-bound (virtual) states,
because $\sigma^{\mathrm{react}}(k)$ does not depend on
$\mathrm{sgn}(\kappa)$, see Eq.~(\ref{eq:sigma_in-full}).

\subsubsection{The elastic cross section}

For the sake of completeness, we also give the expression of the
elastic cross section. Employing again the parametrization
$\tilde{S}=e^{-2i\tilde\alpha k}e^{-2\tilde{\beta}k}$ in
Eq.~(\ref{eq:Sii}), which is then substituted in 
Eq.~(\ref{eq:sigma-el-S-matrix}), we obtain
\begin{eqnarray*}
\lefteqn{
\sigma^{\rm elast} = \frac{\pi}{k^2}
   \frac{2 e^{-2k\tilde{\beta}}}{|p-k|^2}
      \left[ (k^2+|p|^2) \cosh\,2k\tilde{\beta} 
     + 2 k \kappa \sin\,2k\tilde{\alpha} \right.} 
\\
  & & \left. \makebox[0.9cm]{} + (k^2-|p|^2) \cos\,2k\tilde{\alpha} 
         + 2 k q \sinh\,2k\tilde{\beta}
       \right].
\label{eq:sigma-el-full}
\end{eqnarray*} 
Under the assumption of low reactivity, $\tilde{\beta}\to0$,
which is a requirement for having pronounced NTRs, we obtain
\[
\sigma^{\rm elast} = 4\pi|a|^2    \left|\frac{p} {p-k}\right|^2 ,
\]
where $a=\alpha-i\beta$ is the full scattering length.  Although the
expressions for the elastic cross section given above contain four
parameters ($\tilde\alpha$, $\tilde{\beta}$, $q$ and $\kappa$), only
two of them are needed when the resonance pole is very close to the
threshold ($p\approx0$).  Indeed, if the pole contribution is
dominant, we can neglect $\tilde a$, and use $a\approx
a^{\mathrm{res}} = \frac i p$ in the equation above, which we rewrite as
\begin{equation}
\label{eq:sigma-e-q}
\sigma^{\rm elast} =  \frac{4\pi} {(q+k)^2 + \kappa^2}.
\end{equation}
Note the similarity between this result for the elastic cross section
and Eq.~(\ref{eq:sigma-r-q}) for the reactive cross section; thus, in
the limit $k\to0$ we recover again the familiar Wigner behavior,
\[
   \sigma^{\mathrm{elast}} 
    \ \xrightarrow[\ k\rightarrow\,0\ ]\ \ 4\pi |a|^2,
\]
and for the NTR regime we obtain
\[
   \sigma^{\mathrm{elast}} 
    \ \xrightarrow[\ k\,\gg|p|\ ]\ \ \frac{4\pi}{k^2}.
\]


\section{Jost-function approach}
\label{sec:Jost}

We now describe an alternative approach, based on the use of the Jost
function, to obtain the anomalous $k^{-3}$ scaling of the reaction
cross section for the NTR regime described in the previous section.
We shall first give the key concepts for the case of single channel
and short-range interaction $V(R)$, and generalize the results to the
multi-channel case afterward.

\subsection{Single channel case}
      \label{sec:Jost-1}

The regular solution of the radial equation,
\[
\frac{d^2}{dR^2}\phi_k=\left[ 2\mu V(R) - k^2\right] \phi_k(R),
\]
is uniquely specified by initial-value type conditions: $\phi_k=0$ and
$\frac{d}{dR}\phi_k=1$ at $R=0$.  Recall that we are discussing only
the case of s-wave ($\ell=0$) scattering; for higher partial waves,
the initial boundary condition reads: $\phi_{k,\ell}(R)\sim
R^{\ell+1}$.  The asymptotic behavior of the regular solution can be
written as
\begin{equation}
\label{eq:asyAB}
  \phi_k(R) \xrightarrow{\ R\rightarrow \infty\ }
  \  \frac{1}{k}\big[A(k)\sin kR + B(k) \cos kR\big].
\end{equation}
If $k$ is real valued, and assuming $V(R)$ is real, then we can choose
$\phi_k$ to be real; hence, $A(k)$ and $B(k)$ are real valued.
Recasting Eq.~(\ref{eq:asyAB}) in terms of the free solutions
$\exp(\pm ikR)$, we have
\begin{equation}
   \phi_k(R) \xrightarrow{\ R\rightarrow \infty\ }\ \frac{i}{2k}
   \left[ (A-iB) e^{- ikR} - (A+iB)e^{+ikR} \right] ,
\end{equation}
and following the convention used in Refs.~\cite{taylor,newton},
we define the Jost function:
\begin{equation}
\label{eq:Fdef}
   \mathcal F(k) \equiv A(k) - iB(k).
\end{equation}
Conversely, if one introduced $\mathcal F$ in the usual manner, i.e.,
by starting with the asymptotic behavior of $\phi_k$
written as
\begin{equation}
\label{eq:asyF}
   \phi_k(R) \xrightarrow{\ R\rightarrow \infty\ }\ \frac{i}{2k}
   \left[ \mathcal F(k) e^{- ikR} - \mathcal F^*(k)e^{+ikR}  \right],
\end{equation}
one could identify $A=\mathrm{Re}(\mathcal F)$ and
$B=-\mathrm{Im}(\mathcal F)$, provided that both $k$ and $V(R)$ are
real valued.  Note that if one allows for a complex valued $k$, see
\cite{taylor}, a fully general definition of the Jost function can be
obtained by replacing $\mathcal F^*(k)$ with $\mathcal{F}(-k)$ in
Eq.~(\ref{eq:asyF}).  In this article we use real and positive $k$,
unless otherwise specified; when needed, we can allow $k$ to be
complex without any difficulty, because we restrict ourselves to the
case of short-range interactions.  We remark that our analysis of the
$S$~matrix in the previous section (done in the complex $k$ plane) would
be difficult to justify rigorously for the case of long-range
interactions.  However, as we shall see in this section, the approach
based on the Jost function (for real argument $k$) circumvents
this obstacle, because it makes it possible to account for the NTR
phenomenon without referring to the poles of the $S$~matrix in the
complex plane.  Moreover, when it is possible to define the Jost
function for complex $k$, it is easily seen that the two approaches
are equivalent.

Regarding our preference for analyzing $A(k)$ and $B(k)$ instead of
$\mathcal F(k)$ itself, we point out that although the two solutions
$\chi_k^{\pm}(R)\sim e^{\pm ikR}$ are linearly independent in the strict
sense, they will become linearly dependent in the limit $k\rightarrow
0$ (from the {\it numerical} point of view) simply because
$e^{-ikR}\approx e^{ikR}\approx 1$ when $kR\ll 1$.  Thus, it is
desirable to replace $\chi_k^{\pm}$ with a set of solutions which are
more suitable at low $k$, namely $f_k(R)\sim\sin(kR)$ and
$g_k(R)\sim\cos(kR)$.  Since we assume that $\phi_k$, $f_k$, $g_k$ and
$\chi_k^{\pm}$ are exact solutions of the radial equation, the
matching conditions (\ref{eq:asyAB}) and (\ref{eq:asyF}) can be recast
as equalities: $k\phi_k=Af_k+Bg_k$, and $2ik\phi_k=\mathcal
F^*\chi_k^+-\mathcal F\chi_k^-$ respectively.  As is well known, the
Jost function can be expressed as $\mathcal F = W[\chi_k,\phi_k]$,
where $W[u,v]=uv'-u'v$ denotes the Wronskian.  Similarly, for $A$ and
$B$ we have $A=W[g_k,\phi_k]$ and $B=-W[f_k,\phi_k]$, which can be
used in practice to compute $A(k)$ and $B(k)$.

We now shift our attention to the (normalized) physical solution
$\psi_k(R)$, which is regular at $R=0$.  Hence, the physical and
regular solutions only differ by a multiplicative constant, which can
be easily found by comparing the asymptotic boundary condition obeyed by the
physical solution \cite{newton:phi-psi},
\begin{equation}
\label{eq:asypsi}
   \psi_k(R) \xrightarrow{\ R\rightarrow \infty\ }
   \  e^{i\delta} \sin(kR + \delta )
   = \frac i 2 \left(e^{-ikR}-e^{2i\delta}e^{ikR}\right),
\end{equation}
 to the asymptotic behavior in Eq.~(\ref{eq:asyF}).  Indeed, we find
\begin{equation}
   \psi_k(R) = \frac{k}{\mathcal F(k)} \phi_k(R),
   \label{eq:psi-phi}
\end{equation}
and we can also identify the $S$~matrix, $S=e^{2i\delta}=\mathcal
F^*\mathcal F^{-1}$.  We recall the well known relationship between
the phase shift $\delta(k)$ in Eq.~(\ref{eq:asypsi}) and the Jost
function; namely, $\delta = -\arg(\mathcal F)$, which can also be
written as $\tan \delta = B/A$, see Eqs.~(\ref{eq:asyAB},
\ref{eq:Fdef}, \ref{eq:asypsi}).  Also note that if we make use of
$\mathcal F=|\mathcal F|e^{-i\delta}$, then Eq.~(\ref{eq:Fdef}) can be
expressed as $A=|\mathcal F|\cos\delta$ and $B=|\mathcal
F|\sin\delta$.

We emphasize that scattering resonances will affect strongly the $k$
dependence of the normalization constant $k[\mathcal F(k)]^{-1}$ in
Eq.~(\ref{eq:psi-phi}).  Special attention has to be paid to the
denominator $\mathcal F = A-iB$, which could vanish near $k=0$, and
thus produce a near threshold resonance.  Hence, we proceed to analyze
the low $k$ behavior of the Jost function.  Under the assumption of a
short-range potential, it is possible to expand $A(k)$ and $B(k)$ as
power series. Indeed, given that the regular solution, as well as
$\cos(kR)$, are analytic in $k^2$ \cite{newton:page-335}, and that
$\sin(kR)$ is analytic in $k$ and is an odd function of $k$, we can
write:
\begin{eqnarray}
   A(k) & = & A_0 + A_2 k^2 + A_4k^4 + \cdots
\label{eq:A} \\
   B(k) & = & B_1 k + B_3 k^3 + B_5 k^5 + \cdots
\label{eq:B}
\end{eqnarray}
For a finite-range potential, these power series can be rigorously
deduced in a straightforward manner, and their validity can also be
extended to the case of short-range potentials.  However, for
long-range potentials, e.g., $V(R)=-C_n/R^n$ for $R\to\infty$, $A(k)$
and $B(k)$ are no longer analytic functions; their $k$ dependence is
rather complicated, as shown by Willner and Gianturco \cite{gianturco-jost}.
Nevertheless, even in the case of long-range interactions, both $A(k)$
and $B(k)$, and hence $\mathcal F(k)$ are generally well behaved,
albeit nonanalytic; thus, we subscribe to the point of view advocated
by Willner and Gianturco \cite{gianturco-jost}, that the low-$k$
expansion for the Jost function is simpler, more elegant, and more
practical than the corresponding expression for $\tan\delta(k)$ or the
familiar effective range expansion for $k\cot\delta(k)$.  
We argue that one should employ the low-$k$ expansions for $A(k)$ and
$B(k)$ of the Jost function $\mathcal F(k)=A(k)-iB(k)$ rather than low $k$ 
power series expansion of the phase shift (and all other physical 
quantities, such as the cross section). For example,
\[
\tan\delta(k) = \frac{B(k)} {A(k)}
        = \frac {B_1 k + B_3 k^3 + \cdots} {A_0 + A_2 k^2 + \cdots}
\]
should be kept as a fraction, rather than attempting to expand it
consistently to $\mathcal O(k^2)$ or higher.  Indeed, such an
expansion would become unsuitable if $A_0$ were vanishingly small, as
its validity would be restricted to an extremely narrow range, namely
$k^2\ll|A_0/A_2|$, while the expression above remains valid and useful
for a much wider range.  Alternatively, we have
\[
k\cot\delta = \frac  {A_0 + A_2 k^2 + \cdots} {B_1 + B_3 k^2 + \cdots},
\]
which could be expanded as a power series to give the familiar
effective range expansion,
\[
k\cot\delta \approx -\frac{1}{a} + \frac{r_{\mathrm{eff}}} {2} k^2,
\]
with $a=-B_1/A_0$ the scattering length, and
$r_{\mathrm{eff}}=2(A_2/B_1-A_0B_3/B_1^2)=(2/B_1)(A_2+B_3/a)$ the
effective range.  When $A_0$ vanishes, we have $a\rightarrow\pm\infty$
and $r_{\mathrm{eff}}\rightarrow 2A_2/B_1$, and the effective range
expansion remains valid, taking a simplified form: $k\cot\delta
\approx \frac{1}{2} r_{\mathrm{eff}} k^2$.  However, if $B_1$ is
vanishingly small, we have $a\approx 0$ and
$r_{\mathrm{eff}}\rightarrow\pm\infty$, and thus the effective range
expansion is rendered useless.

We now analyze the elastic cross section in terms of the Jost
function:
\[
 \sigma^{\rm elast}(k)  =  \frac{\pi}{k^2} |1-S(k)|^2,
\]
where the S matrix is expressed as
\[
S(k) = e^{2i\delta} = \frac{A+iB}{A-iB}.
\]
Thus, the elastic cross section reads
\begin{equation}
\label{eq:sigma-elast}
     \sigma^{\rm elast}(k)=\frac{4\pi}{k^2} \frac{B^2(k)} {A^2(k)+B^2(k)}.
\end{equation}
In the limit $k\rightarrow 0$, and provided that $A(0)\ne 0$, we recover
the well known Wigner law for s-wave elastic scattering, which is
simply the lowest order approximation,
\[
\sigma^{\mathrm{elast}} \xrightarrow[\;k\rightarrow 0\;] {}
  4\pi\left(\frac{B_1}{A_0}\right)^2 = 4\pi a^2.
\]
The domain of validity for Wigner's threshold law defines the so
called Wigner regime; its extent is governed by the lowest order
coefficients $A_0$ and $B_1$ in the Jost function expansions
(\ref{eq:A}, \ref{eq:B}).  When neither $A_0$ nor $B_1$ is small, the
Wigner regime will cover the entire low energy domain.  Conversely, if
either $A_0$ or $B_1$ is vanishingly small, Wigner's threshold law
will be restricted to a narrow domain near $k=0$, because the higher
order terms in (\ref{eq:A}, \ref{eq:B}) become dominant as $k$
increases, and the $k$-dependence of the cross section in
Eq.~(\ref{eq:sigma-elast}) changes dramatically.

Here, we are especially interested in the case of
vanishingly small $A_0$, which corresponds to a NTR, and we establish
the domain of validity for Wigner's law as follows: in
Eq.~(\ref{eq:sigma-elast}) it is necessary that both $A(k)$ and $B(k)$
be replaced by Eqs.~(\ref{eq:A}, \ref{eq:B}) truncated to the lowest
order, i.e., we require $k\ll|A_0A_2^{-1}|^{1/2}$ and
$k\ll|B_1B_3^{-1}|^{1/2}$.  The former inequality is stronger than the
latter, since $B_1$ is typically finite when $A_0\rightarrow 0$.
Moreover, the $B^2(k)$ term in the denominator of
Eq.~(\ref{eq:sigma-elast}) has to remain negligible; hence, we impose
yet another inequality, $k\ll|A_0B_1^{-1}|=|a|^{-1}$, which is much
stronger than either of the two inequalities above, and thus dictates
the extent of the Wigner regime.

Since the Wigner regime is displaced towards extremely low energies
when $A_0\approx 0$, we need to analyze the behavior of the elastic
cross section for the remainder of the low energy domain.  Thus, we
consider $k$ outside the Wigner regime, but still inside the low
energy domain, such that Eq.~(\ref{eq:B}) can be truncated to the
first term, and Eq.~(\ref{eq:A}) to the second term.  As $k$
increases gradually, $A_0$ becomes negligible compared to both
$A_2k^2$ and $B_1k$.  Moreover, only the latter term survives, as we
have $|B_1|k\gg|A_2|k^2$ throughout the low energy domain.
Consequently, the term $A^2(k)$ in the denominator of
Eq.~(\ref{eq:sigma-elast}) can be neglected entirely, and we
recover the NTR-regime behavior,
\[
\sigma^{\rm elast}(k)\approx\frac{4\pi}{k^2}.
\]

In summary, using the Jost function description, the elastic cross
section can parametrized in a simple way at low energy; namely,
according to our discussion above, we keep only the relevant terms in
Eqs.~(\ref{eq:A}, \ref{eq:B}) to approximate
Eq.~(\ref{eq:sigma-elast}) as
\begin{equation}
\label{eq:sigma-elast-lowE}
\sigma^{\rm elast}(k) = 4\pi \frac{B_1^2} {A_0^2 + B_1^2k^2}
   =4\pi \frac{1}{k^2 +(1/a)^2},
\end{equation}
where $a=-B_1/A_0$ is the full scattering length.  The simple
expression above makes it clear that the the gradual transition from
Wigner's threshold law to the NTR-regime behavior takes place near
$k\approx|a|^{-1}$, where the two terms in the denominator are
comparable.  We remark that Eq.~(\ref{eq:sigma-elast-lowE}) is in
agreement with Eq.~(\ref{eq:sigma-el-kappa}) obtained in the previous
section using the pole factorization of the $S$~matrix, and we shall
discuss their equivalence in Sec.~\ref{sec:equiv}.

\subsection{Multi-channel case}
      \label{sec:Jost-2}

In the case of $N$ coupled channels, the regular solution $\Phi$ is an
$N\times N$ matrix with elements $\phi_{ij}$ satisfying the system of
coupled radial equations,
\begin{equation}
   \frac{d^2}{dR^2} \Phi(\mtrx{k},R) + \mtrx{K}^2 \Phi(\mtrx k,R) 
   = \frac{2\mu}{\hbar^2}{\mtrx V}(R)  \Phi(\mtrx k,R) ,
\end{equation}
with $\mtrx k \equiv ( k_1, k_2, \dots , k_N )$ and $k_n = \sqrt{2\mu
  (E-\varepsilon_n)/\hbar^2}$ the momenta corresponding to channel
threshold energies $\varepsilon_n$.  For simplicity we consider all
channels open, with all $k_n$ real valued and $k_n>0$. $\mtrx K$ is a
diagonal matrix with elements $k_n \delta_{nm}$, while ${\mtrx V}(R)$
is a full matrix containing the couplings $V_{nm}(R)$.  Note that the
diagonal elements of \mtrx V also include centrifugal terms of the
form $\ell_n(\ell_n+1)R^{-2}$.

The regular solution is uniquely specified by initial-value type
conditions at $R=0$,
\[
\phi_{nm}(R)
\ \xrightarrow{\ R\rightarrow\,0\ }\ \delta_{nm}R^{\ell_n+1} .
\]
Assuming that the coupling matrix \mtrx V is real, we  extract two
real valued matrices (\mtrx A and \mtrx B) from the asymptotic
behavior of the regular solution,
\[
\Phi \ \xrightarrow{\ R\rightarrow\,\infty\ }\ \mtrx K^{-\ell-1}
(\mtrx{jA} + \mtrx{nB}),
\]
where $\mtrx K^{-\ell-1}=\mathrm{diag}\{k_n^{-\ell_n-1}\}$, and the
diagonal matrices \mtrx j and \mtrx n contain the free solutions,
i.e., the Riccati--Bessel functions $j_\ell(k_nR)$ and $n_\ell(k_nR)$
for each channel.  Note that, although the asymptotic behavior in the
last equation was written in matrix notation, the coefficients $A_{nm}$
and $B_{nm}$ (i.e., the matrix elements of \mtrx A and \mtrx B) are
extracted independently for each pair of indices $(n,m)$.  Indeed, for
each component ($n$) of any column solution ($m$) contained in $\Phi$,
we write the asymptotic behavior explicitly,
\[
k_n^{\ell_n+1}\phi_{nm}(R) \ \xrightarrow{\ R\rightarrow\,\infty\ }
  \ A_{nm} j_{\ell_n}(k_nR) + B_{nm} n_{\ell_n}(k_nR).
\]
Employing this matching condition for $\phi_{nm}$ (together with a
similar equation for $d\phi_{nm}/dR$) at $R=R_{\mathrm{max}}$ in the
asymptotic region, we  obtain
\begin{eqnarray}
 & & A_{nm} = -k_n^{\ell_n} W[\phi_{nm}, n_{\ell_n}] 
\label{eq:Anm}\\
 & & B_{nm} = +k_n^{\ell_n} W[\phi_{nm}, j_{\ell_n}]
\label{eq:Bnm}
\end{eqnarray}
where $W[\dots]$ stands for the Wronskian.  The Jost matrix is defined
in terms of the matrices \mtrx A and \mtrx B as
\begin{equation}
\label{eq:FAB}
   \mtrx{F} = \mtrx A  - i\, \mtrx B,
\end{equation}
which is a direct generalization of Eq.~(\ref{eq:Fdef}).  The factor
$k_n^{\ell_n+1}$ in the equations above was used for convenience, such
that the Jost matrix is well behaved when $k_n\rightarrow 0$.  In this
paper we restrict our discussion to s-wave scattering, i.e., $\ell=0$
in the entrance channel; note that only the centrifugal term in the
entrance channel is relevant for analyzing the threshold behavior.
The values of $\ell_n$ in all other channels can be arbitrary (as
allowed by the specifics of any given scattering problem).  However,
we set $\ell_n=0$ in all channels to simplify our notation; thus, the
asymptotic behavior of the regular solution reads
\[
k_n\phi_{nm}(R) \ \xrightarrow{\ R\rightarrow\,\infty\ }
  \ A_{nm}\sin(k_nR) + B_{nm}\cos(k_nR),
\]
which is identical to the single channel version, see
Eq.~(\ref{eq:asyAB}).  Equivalently, the asymptotic behavior of $\Phi$
can be written in terms of Jost matrix elements $F_{nm}=A_{nm}-iB_{nm}$
as
\[
 \frac 2 i k_n\phi_{nm}(R) \ \xrightarrow{\ R\rightarrow\,\infty\ }
  \  e^{-ik_nR}  F_{nm}  -  e^{+ik_nR} F^*_{nm} ,
\]
where $F^*_{nm}=A_{nm}+iB_{nm}$ is the complex conjugate of $F_{nm}$.

The normalized (physical) solution $\Psi$ can be written in terms
of the regular solution $\Phi$,
\begin{equation}
   \Psi (R) = \Phi(R)  \mtrx{F}^{-1} \mtrx K ,
   \label{eq:psi-multi}
\end{equation}
and it has the well known asymptotic behavior,
\begin{equation}\label{eq:psi-S}
\psi_{nm}(R) \ \xrightarrow{\ R\rightarrow\,\infty\ }\  \frac{i}{2}
  \left[ e^{-ik_nR}\delta_{nm}
    - e^{+ik_nR} \left(k_n^{-1/2}S_{nm}k_m^{1/2}\right)\right],
\end{equation}
where $S_{nm}$ are the elements of the S matrix,
\begin{equation}
\label{eq:S}
\mtrx S = \mtrx K^{-1/2} (\mtrx{F})^* (\mtrx{F})^{-1} \mtrx K^{1/2}.
\end{equation}
Equations (\ref{eq:psi-multi}) and (\ref{eq:S}) contain the inverse of
the Jost matrix, which we write explicitely as
\begin{equation}
\label{eq:F-inv}
   \mtrx{F}^{-1} = \frac{1}{\det(\mtrx F)} 
   \left[ \mathrm{Cof}(\mtrx F) \right]^T,
\end{equation}
with $[{\rm Cof}(\mtrx F)]^T$ the transpose of the matrix of cofactors
of \mtrx F, and $\det(\mtrx F)$ the determinant of \mtrx F.

As in the single-channel case, we focus our attention on the
denominator, i.e., the determinant of the Jost matrix.  We are
interested in the situation when $\det(\mtrx F)$ is vanishingly small,
such that the scattering cross sections are resonantly enhanced;  if
$T_{nm}$ are the elements of the T matrix, $\mtrx T=\mtrx 1-\mtrx S$,
we have
\[
\sigma(n\leftarrow m)  \sim  \frac{\pi}{k_m^2}|T_{nm}|^2
   \sim |\det(\mtrx F)|^{-2}.
\]
Although the determinant of the Jost matrix cannot vanish on the real
axis \cite{newton:page:xxx}, it can \emph{almost} reach zero.  Also,
recall that we only consider the case of a scattering problem without
closed channels, thus eliminating the possibility of Feshbach
resonances.  Therefore, by extrapolation from the single-channel case,
the only remaining possibility is that of a potential resonance at low
energy; such a resonance would correspond to a quasi-bound state
near the threshold of the channel with the highest energy
asymptote, which we take as our entrance channel ($n=1$).  For
clarity, we assume
$E>\varepsilon_1>\varepsilon_2>\cdots>\varepsilon_n>\cdots>\varepsilon_N$.

When $\det(\mtrx F)\approx 0$ at a channel threshold, the factor
$|\det(\mtrx F)|^{-2}$ in the equation above will be responsible for a
resonance enhancement, and will also affect the threshold behavior of
cross sections.  We now give a detailed analysis of a NTR in the
entrance channel, $n=1$, and we start by using a cofactor expansion of
the determinant along the first row,
\begin{equation}
   \det(\mtrx{F}) = \sum_{m=1}^N F_{1m}C_{1m} ,
   \label{eq:det}
\end{equation}   
where $C_{1m}$ is the cofactor corresponding to the element $(1,m)$.
Next, we extract $C_{11}$ outside the sum, and we isolate the first
term (i.e., the diagonal element $F_{11}$) to recast $\det(\mtrx F)$
as
\begin{equation}
   \det(\mtrx F) = C_{11} \left(F_{11} + f_{11} \right)  ,
    \label{eq:det-C11}
\end{equation} 
where
\begin{equation}
\label{eq:f11}
 f_{11} \equiv  \frac{1}{C_{11}} \sum_{m\ne 1} F_{1m} C_{1m} .
\end{equation}
Separating the real and imaginary parts of the last factor in
Eq.~(\ref{eq:det-C11}), namely $F_{11}+f_{11}=\mathcal A -i\mathcal B$,
and defining
\[
   D \equiv |F_{11}+f_{11}|^2 = \mathcal A^2 +\mathcal B^2,
\]
we can write
\begin{equation}
\label{eq:det-D}
|\det(\mtrx F)|^2 = |C_{11}|^2|\mathcal A - i\mathcal B|^2 = |C_{11}|^2 D.
\end{equation}

Under the assumption that $f_{11}$ is suitably small, and provided
that $C_{11}\not\approx0$, only the factor
$D\approx|F_{11}|^2=A^2_{11}+B^2_{11}$ can be responsible for NTRs.
In order to clarify these assumptions, we recall that, as we explained
in Sec.~\ref{sec:Smat}, due to the coupling of the entrance channel
with other open channels, the resonance pole will be pushed away from
the threshold; thus, in the limit of strong coupling, the existence of
a near threshold resonance will become highly unlikely.  Conversely,
the appearance of a prominent NTR  requires the entrance channel be
weakly coupled to all other open channels.  To simplify, one can use a
coupling strength parameter $\lambda$ to redefine these particular
couplings ($V_{1n}=V_{n1}\to\lambda V_{n1}$) which will become
vanishingly small when $\lambda\to 0$.  Hence, we assume the regular
solution matrix $\Phi$ is \emph{almost} block diagonal, with a small
$1\times 1$ block containing $\phi_{11}$ and a large
$(N-1)\times(N-1)$ block corresponding to all other open channels.
The remaining elements, i.e., $\phi_{1m}$ with $m\ne1$ in the first
row, and $\phi_{n1}$ with $n\ne1$ in the first column are vanishingly
small when $\lambda\to 0$.  Consequently, from Eqs.~(\ref{eq:Anm},
\ref{eq:Bnm}, \ref{eq:FAB}) we have (for $n\not=1$) $F_{n1}=\mathcal
O(\lambda)$ when $\lambda\to0$, and thus $C_{1m}=\mathcal{O}(\lambda)$
for $m\not=1$.  Note that $C_{11}$ is independent of $\lambda$ in
zeroth order, and hence we assume $C_{11}=\mathcal O(1)$.  Finally, we
also have $F_{1m}=\mathcal O(\lambda)$ for $m\not=1$, and from
Eq.~(\ref{eq:f11}) we have $f_{11}=\mathcal{O}(\lambda^2)$ when
$\lambda\to0$.

Regarding the caveat of an accidental vanishing of $C_{11}$ near the
threshold of the entrance channel, recall that in the limit
$\lambda\to0$ the channel $n=1$ becomes decoupled from the remaining
$N-1$ channels, and we have $\lim_{\lambda\to0}C_{11}=\det\big[\mtrx
  F^{(N-1)}\big]$, where $\mtrx F^{(N-1)}$ is the Jost matrix for the
scattering problem with $N-1$ channels.  Thus, $\det\big[\mtrx
  F^{(N-1)}\big]$ cannot be vanishingly small for collision energies
$E>\varepsilon_1$, since $E$ is very high above $\varepsilon_2$ (which
is the highest threshold for the scattering problem with $N-1$
channels).  Indeed, as we assume $\varepsilon_1-\varepsilon_2$ is much
larger than the generic low-energy scale of the scattering problem;
thus, it is safe to factor out $C_{11}=\mathcal O(1)$ as a background
contribution, as we already anticipated in the equations above.

Denoting $k=k_1$, we now focus on the $k$ dependence of $D(k)$ in
Eq.~(\ref{eq:det-D}), which stems from the behavior of $F_{11}(k)$ and
$f_{11}(k)$.  We emphasize that all quantities are analyzed as
functions of the single variable $k$.   All other momenta, i.e.,
$k_n$ for channels $n=2,3,\ldots,N$, can be easily expressed in terms
of the entrance channel momentum:  using $k_n^2=2\mu(E-\varepsilon_n)$,
we write
\[
  k_n   =  \sqrt{\Delta_n + k^2},
\]
with $\Delta_n = 2\mu (\varepsilon_1-\varepsilon_n)$.  In the
ultracold limit ($k\to0$) we have $k^2\ll\Delta_n$ for all $n\ne1$,
and we use a power series expansion:
\[
  k_n(k^2) \approx \sqrt{\Delta_n} + \mathcal O(k^2).
\]
Moreover, the regular solution $\Phi$ is a function of $k^2$ (rather
than $k$ itself), and we deduce from Eqs.~(\ref{eq:Anm}, \ref{eq:Bnm})
that except for the first row, all matrix elements $A_{nm}$ and
$B_{nm}$  are also functions of $k^2$.  Thus,
when $k\to0$ we  have
\[
F_{nm}(k^2)  =  F_{nm}(0) + \mathcal O(k^2), \quad\mathrm{for}\ n\ne1,
\]
Similarly,  the cofactors corresponding to the first row of \mtrx F
can be written as
\[
C_{1m}(k^2) = C_{1m}(0) +\mathcal O(k^2), \quad\mathrm{for\ all}\ m,
\]
which ensures that the ratios $c_m\equiv C_{1m}/C_{11}$ in
Eq.~(\ref{eq:f11}) can be expanded in a similar fashion:
\[
c_m(k^2) = c_m(0) + \mathcal O(k^2).
\]
Rewriting Eq.~({\ref{eq:f11}) as
\begin{equation}
\label{eq:f11cm}
f_{11} = \sum_{m\ne1} F_{1m} c_{m},
\end{equation}
and separating its real and imaginary parts,
we have
\begin{equation}
\label{eq:Ref11}
 \mathrm{Re}(f_{11}) = \sum_{m\not=1}\big[A_{1m}(k)\mathrm{Re}c_m(k)
  +B_{1m}(k)\mathrm{Im}c_m(k)\big],
\end{equation}
\begin{equation}
\label{eq:Imf11}
 \mathrm{Im}(f_{11}) = \sum_{m\not=1}\big[A_{1m}(k)\mathrm{Im}c_m(k)
  -B_{1m}(k)\mathrm{Re}c_m(k)\big].
\end{equation}
In these equations, special attention has to be paid to
$A_{1m}(k)$ and $B_{1m}(k)$.  From Eqs.~(\ref{eq:Anm},
\ref{eq:Bnm}) we obtain the expansions:
\begin{eqnarray}
\label{eq:A1m}
A_{1m}(k) & = & A_{1m}(0)  + k^2A_{1m}^{(2)} + \mathcal O(k^4),
\\
\label{eq:B1m}
B_{1m}(k) & = & \qquad kB_{1m}^{(1)} +k^3B_{1m}^{(3)} + \mathcal O(k^5).
\end{eqnarray}
Since the series of $B_{1m}(k)$ contains odd powers, and $c_m(k)$ in
Eq.~(\ref{eq:f11cm}) are complex quantities, the power series in
Eqs.~(\ref{eq:Ref11}, \ref{eq:Imf11}) contain both odd and even
powers of $k$,
\begin{eqnarray}
\mathrm{Re}(f_{11})  & = & a(0) + ka^{(1)} + k^2a^{(2)} + \mathcal O(k^3),
\\
\mathrm{Im}(f_{11})  & = & b(0) + kb^{(1)} + k^2b^{(2)} + \mathcal O(k^3).
\end{eqnarray}
Thus, the power series of $\mathcal A(k)$ and $\mathcal B(k)$ will also
contain all possible terms, i.e., both odd and even powers:
\begin{eqnarray}
\label{eq:Afull}
  \mathcal A(k) & = & \mathcal A_0 + k\mathcal A_1 + k^2\mathcal A_2 +\cdots,
\\
\label{eq:Bfull}
  \mathcal B(k) & = & \mathcal B_0 + k\mathcal B_1 + k^2\mathcal B_2 +\cdots
\end{eqnarray}
The coefficients $\mathcal A_\nu$ and $\mathcal B_\nu$ will be
expressed in terms of $A_{1m}^{(\nu)}$, $B_{1m}^{(\nu)}$, $a^{(\nu)}$
and $b^{(\nu)}$.  Indeed, recall that $\mathcal A=\mathrm{Re}
(F_{11}+f_{11})=A_{11}+\mathrm{Re}(f_{11})$ and
$\mathcal{B}=-\mathrm{Im}(F_{11}+f_{11})=B_{11}-\mathrm{Im}(f_{11})$.
From the equations above, we obtain for the coefficients up to order
$\nu=2$,
\begin{eqnarray*}
\mathcal A_0 = \mathcal A(0)  & = &  a(0) + A_{11}(0)
\\
\mathcal A_1 & = &  a^{(1)}
\\
\mathcal A_2 & = &   a^{(2)} + A_{11}^{(2)}
\end{eqnarray*}
and
\begin{eqnarray*}
\mathcal B_0 = \mathcal B(0) & = & -b(0)
\\
\mathcal B_1 & = &  - b^{(1)} + B_{11}^{(1)}
\\
\mathcal B_2 & = &   -b^{(2)}.
\end{eqnarray*}
Under the assumption of weak coupling ($\lambda\to0$), we have
$f_{11}=\mathcal O(\lambda^2)\to0$.  Thus, all $a^{(\nu)}$ and
$b^{(\nu)}$ are negligibly small; however, their presence in the
equations above is necessary, because they limit the effects caused by
 near threshold resonances.  Indeed, note that $\mathcal A_0$ and
$\mathcal B_0$ cannot vanish simultaneously; in other words, one
cannot have: $A_{11}(0)=0$ and $a(0)=0$ and $b(0)=0$.  Were that the
case, $D(k)$ and hence $\det(\mtrx F)$ would vanish at $k=0$, which is
impossible (as we mentioned earlier).  Nevertheless, when $\lambda$ is
small, so are $a(0)$ and $b(0)$.  Also, as in the single-channel case,
it is possible that $A_{11}(0)\approx0$.  Thus, although the quantity
$D(k=0)=\mathcal A_0^2+\mathcal B_0^2$ cannot vanish, it could reach
very small values (which is the signature of NTR).

Finally, we analyze the threshold behavior of cross sections in the
presence of a NTR; from Eqs.~(\ref{eq:S}, \ref{eq:F-inv}) we obtain
\[
S_{nm} = \delta_{nm}  +  \sqrt{\frac{k_m}{k_n}}
  \frac{\;2i\sum_j B_{nj}C_{mj}}{\det(\mtrx F)}.
\]
We only consider scattering from the ultracold channel $m=1$ to all
final channels $n=1,2,3,\ldots,N$; thus, using the notation $k=k_1$,
the cross sections read
\[
\sigma(n\leftarrow 1) \sim\frac{\pi}{k^2} |T_{n1}(k)|^2
   \sim \frac{|\sum_jB_{nj}C_{1j}|^2}{kk_n|C_{11}|^2D(k)}.
\]
Except for $D(k)$, all other quantities are finite when $k\to0$, and
can be replaced by their values at $k=0$, i.e., $B_{nj}(0)$,
$C_{1j}(0)$, $C_{11}(0)$ and $k_n(0)$, which amount to an overall ($n$
dependent) constant.  This allows us to simplify the expression above;
for reactive (or inelastic) scattering, $n\not=1$, we obtain
\begin{equation}
\label{eq:sigma-r-D}
\sigma^{\mathrm{react}}(n\leftarrow 1) \sim
   \frac{\mathrm{const}(n)}{D(k)}  k^{-1},
\end{equation} 
while for elastic scattering, we have $k_n=k$ and $B_{1j}(k)\sim
k$, and the cross section reads
\begin{equation}
\label{eq:sigma-e-D}
\sigma^{\mathrm{elast}}(1\leftarrow 1) \sim \frac{\mathrm{const}'}{D(k)}.
\end{equation}
Recall  $D(k)=\mathcal A^2(k)+\mathcal B^2(k)$, which is expanded at
low $k$,
\begin{equation}
\label{eq:D}
D(k) = D_0 + kD_1 +k^2D_2 + \cdots,
\end{equation}
with
\begin{eqnarray*}
D_0=D(0) & = &  [b(0)]^2 + [a(0)+A_{11}(0)]^2 ,
\\
D_1 & = & 2[-b(0)\mathcal B_1 + a^{(1)}\mathcal A_0],
\\
D_2 & = & \mathcal B_1^2 + [a^{(1)}]^2 + 2[\mathcal A_0\mathcal A_2+b(0)b^{(2)}].
\end{eqnarray*}
As argued above, $D_0$ cannot vanish identically; however, $D_0$ is
small in the case of a NTR, and hence it is the most important
parameter characterizing the NTR.  Note that the first order term in
Eq.~(\ref{eq:D}) can be neglected when both the weak coupling
($\lambda\to0$) assumption and the NTR assumption ($A_{11}(0)\approx
0$) are valid; indeed, we have $D_1\approx-2b(0)B_{11}^{(1)}=\mathcal
O(\lambda)$.  Thus, even for a pronounced NTR, i.e., when $D_0$
approaches its minimum value, $D_0\approx D_{\mathrm{min}}\approx
[b(0)]^2=\mathcal O(\lambda^2)$, the linear term will become
negligible compared to $D_0$ for $k\ll
D_0/|D_1|\approx|b(0)/B_{11}^{(1)}|$.  In the ultracold limit
($k\to0$) the quadratic term in Eq.~(\ref{eq:D}) can be neglected
as well; specifically, the requirement is $k\ll\sqrt{D_0/D_2}$.
Making use of the approximations $D_0\approx[b(0)]^2$ and
$D_2\approx[B_{11}^{(1)}]^2$, we have
$\sqrt{D_0/D_2}\approx|b(0)/B_{11}^{(1)}|$.  Hence, the strong
inequality $k\ll\sqrt{D_0/D_2}$ defines the Wigner regime, since we
can ignore both the linear and quadratic terms in
Eq.~(\ref{eq:D}).  Indeed, we have $D(k)\approx D_0$, and we
recover Wigner's threshold laws: $\sigma^{\mathrm{react}}\sim k^{-1}$
and $\sigma^{\mathrm{elast}}\approx$ constant.

The linear term in Eq.~(\ref{eq:D}) can also be neglected outside
the Wigner regime, i.e., for the remainder of the low $k$ domain;
indeed, when $k\gg|b(0)/B_{11}^{(1)}|$, we have $k^2D_2\gg kD_1$ and
also $k^2D_2\gg D_0$, which ensures that $D(k)\approx k^2D_2$.  Thus,
discarding the linear term altogether in Eq.~(\ref{eq:D}), we
simplify Eqs.~(\ref{eq:sigma-r-D}) and (\ref{eq:sigma-e-D}),
and we obtain
\begin{equation}
\label{eq:sigma-D-simple}
k^{0,1}\sigma^{\mathrm{elast,\ react}} \sim \frac {1} {D_0+k^2D_2},
\end{equation}
where the prefactors $k^0=1$ and $k$ correspond to elastic and
reactive scattering, respectively.  We emphasize that this expression
is a good approximation throughout the entire low-$k$ regime, and
despite its simplicity, it accounts for the combined Wigner and
NTR-regime behavior.  For reactive scattering we have
\[
\sigma^{\mathrm{react}} \sim \left\{
\begin{array}{l}
k^{-1},\quad\mathrm{Wigner\ regime\ for}\  k\ll\sqrt{D_0/D_2}
\\
k^{-3},\quad\mathrm{NTR\ regime\ for}\  k\gg\sqrt{D_0/D_2},
\end{array}
\right.
\]
while for elastic scattering,
\[
\sigma^{\mathrm{elast}} \sim \left\{
\begin{array}{ll}
k^0,  & \quad\mathrm{Wigner\ regime}
\\
k^{-2},  & \quad\mathrm{NTR\ regime,}
\end{array}
\right.
\]
According to Eq.~(\ref{eq:sigma-D-simple}) there is a smooth
transition between the two types of behavior (Wigner and NTR) which
takes place around $k=\sqrt{D_0/D_2}$, where all three terms in
Eq.~(\ref{eq:D}) are comparable.  Thus, the linear term in
Eq.~(\ref{eq:D}) only plays a minor role near the transition, while
becoming negligible for both Wigner and NTR regimes.

\begin{figure}[h]
\includegraphics[clip,width=\linewidth]{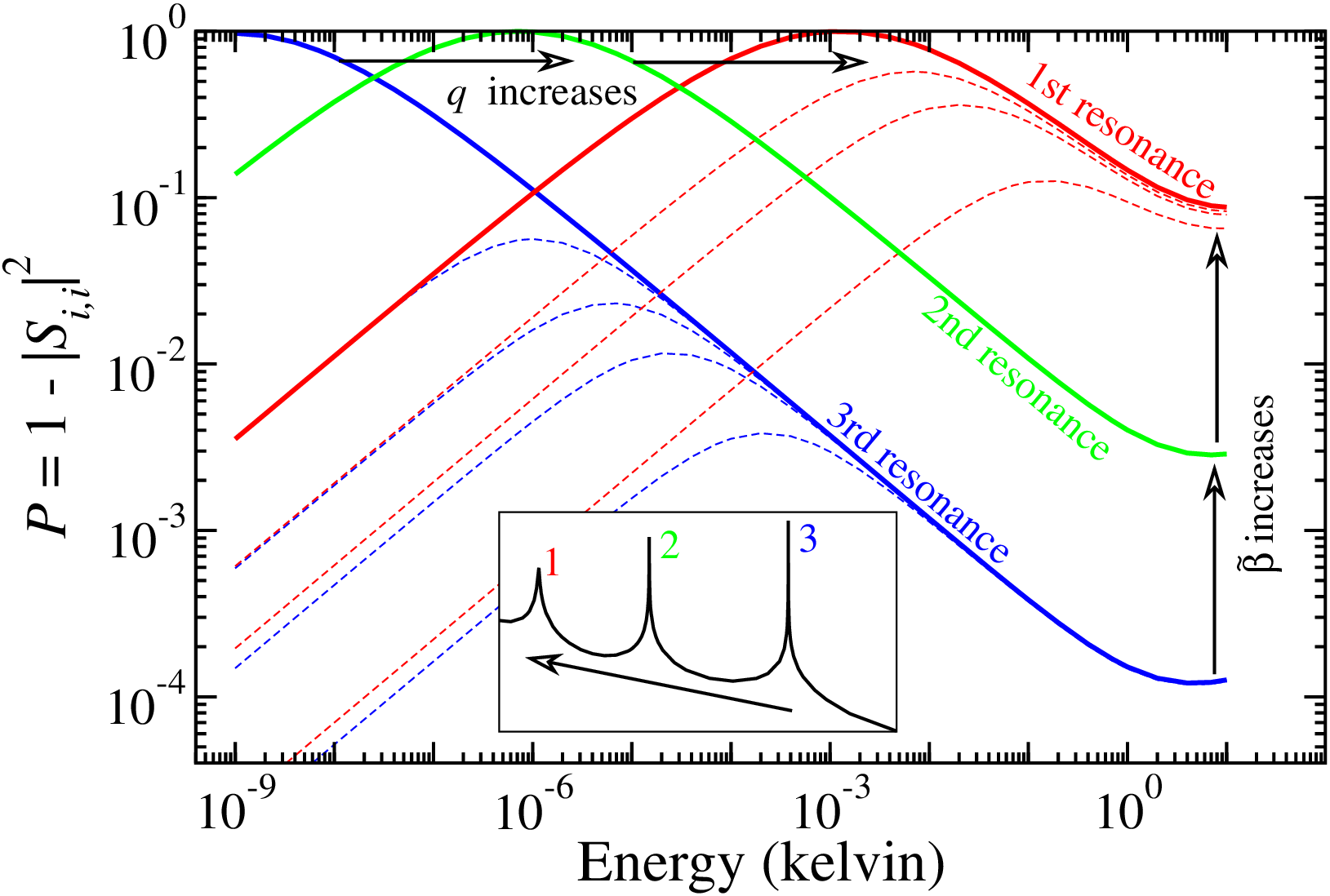}
\caption{\label{fig:trend} Comparison of reaction probabilities for
  the resonances of H$_2$ + Cl.  The imaginary part of the scattering
  length ($\beta$, see Fig.~\ref{fig:alpha-beta}) is shown in the
  inset to help identify the resonances.  The three solid curves
  correspond to the three critical values of $m$ (for the first,
  second and third resonance, which are shown in red, green and blue,
  respectively).  The dashed curves correspond to values of $m$ near
  the first and third resonance.  The arrows indicate the increase of
  reactivity, and the increase of $\tilde\beta$ and $q$, as $m$
  decreases.}
\end{figure}

We remark that when the coupling strength $\lambda$ increases, and the
system becomes more reactive, $a^{(\nu)}$ and $b^{(\nu)}$ will
increase; hence, $D_0$ and $D_1$ will become large (even if
$A_0\approx0$ still holds) and the NTR regime will be pushed to higher
values of $k$, where higher-order terms will become dominant.
Eventually, the signature of the NTR will disappear, as suggested in
Fig.~\ref{fig:trend}, which shows that for the lowest reactivity
$\tilde\beta$ corresponding to the third resonance, the pole is
located at very low $k$, and the NTR regime extends to a sizable range
for larger $k$.  The extent of the NTR regime is smaller for the
second resonance corresponding to a larger $\tilde\beta$, and smaller
still for the first resonance with the largest reactivity
$\tilde\beta$.  Thus, in the limit of strong coupling, the very
existence of a quasi-bound state of the atom--dimer complex becomes
unlikely, since such a complex will be short lived; consequently, NTRs
cannot exist (or are less pronounced) for highly reactive systems.

\section{Equivalence between different approaches}

\label{sec:equiv}

In the case of purely elastic scattering (single channel), the
equivalence of the two approaches is based on the simple relationship
between the $S$ matrix and the Jost function, see
Eq.~(\ref{eq:Sjost}).  The expressions (\ref{eq:Sfact}) and
(\ref{eq:A}, \ref{eq:B}) merely give different parametrizations
for the $S$ matrix, and hence for the elastic cross section.  To
elucidate their equivalence, we first note that the pole factorization
(\ref{eq:Sfact}) relies on the power series expansion of the Jost
function around its zero (at $k=p$).  Recall that we assume a short
range potential, such that $\mathcal F(k)$ is analytic in the entire
complex $k$ plane, or at least inside a wide region containing the
threshold $k=0$.  We thus write
\[
\mathcal F(k) = \sum_{n=1}^{\infty} F_n(p-k)^n,
\]
where the coefficients $F_n$ are complex numbers.  The absence of the
term for $n=0$ in the sum above makes it explicit that $\mathcal F(p)=0$,
and allows for the factorization $\mathcal F(k)=(p-k)\tilde F(k)$.
The background contribution $\tilde F(k)$ is also analytic,
and assuming it to be slowly varying within a small domain containing
both $k=p$ and $k=0$, we can truncate its power series in the low
energy domain:
\[
\tilde F(k) \approx \tilde F_0 + \tilde F_1 k. 
\]
Note that for $\tilde F(k)$ we employ a power series expansion around
$k=0$, which is more convenient.  Hence, the $S$ matrix reads
\[
S(k) = \frac{\mathcal F(-k)}{\mathcal F(k)}
     = \frac{p+k}{p-k} \frac{\tilde F(-k)}{\tilde F(k)},
\]
and we recognize the background contribution, see Eq.~(\ref{eq:Sfact}),
\[
\tilde S(k) = \frac{\tilde F(-k)}{\tilde F(k)},
\]
which is slowly varying near $k=0$, due to the smooth behavior of
$\tilde F(k)$.  Truncating the power series of $\tilde S$ to first
order, we have $\tilde S(k) \approx 1 - 2i\tilde a k$.  The unitarity
of $\tilde S$ on the real axis ensures that the coefficient of the
first order term in this expansion is purely imaginary, and we
have used the customary notation, namely the background scattering
length $\tilde a$.  Thus, in the pole factorization approach, the full
$S$ matrix at low $k$ reads
\[
S(k) \approx \frac{p+k}{p-k} (1 - 2i\tilde a k).
\]

We now compare this result with the equivalent expression obtained in
the explicit Jost function approach, namely in terms of $\mathcal
F(k)= A(k)-iB(k)$.  We emphasize that $A(k)$ and $B(k)$ are expanded
in power series around $k=0$ only, and we remark that for a short
range potential, the power series of $A(k)$ and $B(k)$ remain valid
when $k$ takes complex values; moreover, the expansion coefficients
$A_n$ and $B_n$ in Eqs.~(\ref{eq:A}, \ref{eq:B}) remain real numbers.
Therefore, if a zero of the Jost function (i.e., a pole of the $S$
matrix) is located on the imaginary axis, $p=i\kappa$, then $\kappa$
is a real valued solution of the equation
\[
A_0 + B_1\kappa - A_2\kappa^2 - B_3\kappa^3 + \cdots = 0,
\]
with all coefficients  real.  In the case of a NTR, $A_0$ is
vanishingly small, and to a good approximation we have
\[
\kappa\approx - \frac{A_0}{B_1} = \frac 1 a ,
\]
where $a$ is the full scattering length, which agrees with
Eq.~(\ref{eq:a-kappa}).  As we discussed in Sec.~\ref{sec:Jost-1}, the
$S$~matrix is simply written as 
\[
S(k) = \frac{A_0 + iB_1k + A_2k^2 + \cdots}{A_0 - iB_1k + A_2k^2 - \cdots}.
\]
When the resonance pole $p=i\kappa$ is very close to $k=0$, this
expression truncated to second order becomes equivalent with the
similar expression from the pole factorization approach, which has the
advantage of using the exact value of the resonance pole (extracted as
a fitting parameter from the computed scattering cross section, as it
was done in Sec.~\ref{sec:Smat}).  However, for complex $k$, the very
definition of the $S$ matrix becomes problematic, as it imposes strict
requirements for the asymptotic behavior of the interaction potential.
Hence, the direct approach based on Jost functions for real $k$ has
the advantage of being fully general.  We also emphasize that the Jost
function has a smooth $k$-dependence, and it can be easily
interpolated on a rather coarse $k$-mesh, which is very convenient in
practice; nevertheless, this advantage has remained largely
overlooked, despite being recognized more than four decades ago by
Heller and Reinhardt \cite{rick.heller}.

Regarding the coupled-channel case, we remark that
Eq.~(\ref{eq:sigma-D-simple}) obtained in the Jost function approach
represents a parametrization which is identical to
Eq.~(\ref{eq:sigma-r-q}) from the pole factorization treatment,
provided we simplify the latter to read $k\sigma^{\rm react}(k)=4\pi
q(k^2+|p|^2)^{-2}$.  We can thus identify $D_0/D_2=|p|^2=q^2+\kappa^2$
in the denominator, and we regard $q$ in the numerator as an overall
measure of background reactivity (due to the couplings $V_{n1}=V_{1n}$
between the entrance channel and all other channels).

\section{Qualitative interpretation based on the wave function amplitude}
\label{sec:wave}

A very simple picture emerges from the treatment using the Jost
function; indeed, as noted for the single-channel discussion in
Sec.~\ref{sec:Jost-1}, the amplitude of the physical wave function is
modulated by the inverse of the Jost function ${\mathcal F}^{-1}(k)$,
see Eq.~(\ref{eq:psi-phi}).  This scaling property is especially
useful in the low energy domain, i.e., for $k\lessapprox|\tilde{a}|^{-1}$,
when the regular solution in the \emph{short range}
($R\lessapprox|\tilde a|$) region is nearly $k$-independent:
$\phi_k(R)\approx\phi_0(R)$.  Hence, within the short-range region,
Eq.~(\ref{eq:psi-phi}) reads $\psi_k(R)\approx k{\mathcal
  F}^{-1}(k)\phi_0(R)$, and the $k$~dependence of the physical
solution is driven entirely by the scaling factor $k{\mathcal
  F}^{-1}(k)$.

In the absence of NTRs, the Jost function can also be replaced by its
value at $k=0$, ${\mathcal F}(0)=A_0$, and we obtain the well known
linear scaling of the short-range amplitude of the physical
wave function, $\psi_k(R)\approx kA_0^{-1}\phi_0(R)$, characteristic to
the Winger regime.  However, when ${\mathcal F}(0)=A_0$ is vanishingly
small, it produces a resonance enhancement for $\psi_k(R)$, and also
the anomalous NTR behavior,
\begin{equation}\label{eq:psi-scaling}
\left|\psi_k(R)\right|^2 \approx \frac{k^2}{A_0^2 + B_1^2k^2}
\left|\phi_k(R)\right|^2.
\end{equation}
We remark that this simple expression for the single-channel case
already contains the main result, Eq.~(\ref{eq:sigma-D-simple}),
obtained rigorously in Sec.~\ref{sec:Jost-2} for the coupled-channel
case.  This can be justified qualitatively, if we rely on the
following physical argument; under the assumption of small couplings,
the entrance-channel component ($\psi_1$) of the physical wave function
follows the same scaling as in the single-channel case, see
Eq.~(\ref{eq:psi-multi}).  This $k$-dependence in the equation above
will be imprinted (via the couplings $V_{nm}$) on all the other
components $\psi_n$.  Although the entrance-channel component obeys
Eq.~(\ref{eq:psi-scaling}) only at short range, all other components
will obey it for all $R$ (including the long-range region).  Thus,
according to Eq.~(\ref{eq:psi-S}) for $m=1$, we extract the matrix
elements $S_{n1}$ from the asymptotic behavior of $\psi_n(R)$, and we
obtain for $n\ne1$
\[
\frac{k}{k_n} \left|S_{n1}\right|^2 \propto\left|\psi_n\right|^2
\propto \frac{k^2}{A_0^2 + B_1^2k^2},
\]
where we used $k=k_1$ for the entrance channel, and 
$k_n(k)\approx\sqrt{\Delta_n}$ is constant.  
Finally, we obtain for the reaction cross section,
\[
 k \sigma^{\rm react}(n\leftarrow 1) \propto \frac{1}{A_0^2 + B_1^2k^2},
\]
which is the same result obtained in Eq.~(\ref{eq:sigma-D-simple}),
with $D_0\approx A_0^2$ and $D_2\approx B_1^2$.

\section{Conclusions}
\label{sec:conclusion}

In conclusion, we found that a near threshold resonance will affect
dramatically the behavior of the $s$-wave cross section at low energy.
In fact, the low energy domain will be divided in two distinct regimes:
the Wigner regime with the well known scaling ($\sigma^{\rm
  react}\propto k^{-1}$) and the NTR regime with the anomalous scaling
$\sigma^{\rm react}\propto k^{-3}$.  We derived simple analytical
expressions for both elastic and reaction (inelastic) cross sections,
which depend on the position of the pole of the $S$ matrix, and found
very good agreement with results obtained numerically for the full
reactive scattering problem.  We explained the anomalous NTR-regime
behavior using two different approaches---one based on the pole
factorization of the $S$ matrix, and a more general treatment based on
the low energy expansion of the Jost function.

We remark that the $k^{-3}$ NTR-regime behavior (for s-wave) is a
general feature, and it can be a rather common occurrence in
scattering problems; however, unless the resonance pole is very close
to $k=0$, the presence of the NTR can be somewhat inconspicuous, or it
can be masked by higher partial wave contributions.  Although previous
work hinted at such anomalous resonant behavior---which was explored
either by using mass-scaling \cite{Bodo:JPB:2004,Alex:PRA2010:He-H2},
or external fields \cite{JMH:PRL:2009:Feshbach}---the new NTR-regime
behavior shown in Eq.~(\ref{eq:sigma_in-full}) was discussed only
recently \cite{our-NTR}. These studies explored the effects of
mass-scaling or magnetic field scanning at fixed scattering energy,
e.g., see \cite{NJP-Hutson-2007}. In our work, we revealed the effect
of NTRs in two benchmark atom-diatom reactive scattering systems by
mass-scaling (H$_2$+Cl and H$_2$+F), but it also appears in any system
with a zero-energy resonance, such as in photoassociation at ultralow
temperatures \cite{Crubellier:JPB:2006,FOPA}, ultracold collisions in
general \cite{Cs2:resonance}, or spin-relaxation in ultracold atomic
samples \cite{Cs2:spin}. The modified ($k^{-3}$) behavior of the
reaction (inelastic) cross section will play a role in the
interpretation of experiments such as \cite{narevicius-2012}, and also
for theories  based on Wigner's threshold law developed
to account for resonances in ultracold molecular systems
\cite{Bohn:PRA87:2013}.

This work was partially supported by the US Department of Energy,
Office of Basic Energy Sciences (RC), and the Army Research
Office Chemistry Division (I. S.).


\begin{thebibliography}{99}
\bibitem{RMP-FR}
  C. Chin, R. Grimm, P. Julienne, and E. Tiesinga,
  Rev. Mod. Phys. {\bf 82}, 1225 (2010).
\bibitem{paper-JILA}
  M. H. G. de Miranda, A. Chotia, B. Neyenhuis, D. Wang, G.~Qu\'em\'ener,
  S. Ospelkaus, J. L. Bohn, J. Ye, and D. S. Jin,
  Nature Phys. {\bf 7}, 502 (2011).
\bibitem{Jason-PRL}
  J. N. Byrd, J. A. Montgomery, Jr., and R. C\^ot\'e, 
  Phys. Rev. Lett. {\bf 109}, 083003 (2012).
\bibitem{RMP-bose}
  F. Dalfovo, S. Giorgini, L. P. Pitaevskii, and S. Stringari,
  Rev. Mod. Phys. \textbf{71}, 463 (1999);
  A. J. Leggett,
  Rev. Mod. Phys. \textbf{73}, 307 (2001).
\bibitem{RMP-fermi}
  S. Giorgini, L. P. Pitaevskii, and S. Stringari,
  Rev. Mod. Phys. \textbf{80}, 1215 (2008).
\bibitem{efimov}
  T. Kraemer, M. Mark, P. Waldburger, J. G. Danzl, C. Chin, B. Engeser,
  A. D. Lange,  K. Pilch, A. Jaakkola, H.-C. N\"agerl, and R. Grimm,
  Nature {\bf 440}, 315 (2006).
\bibitem{carr2009} 
  L. Carr, D. DeMille, R. Krems, and J. Ye,
  New J. Phys. {\bf 11}, 055049 (2009).
\bibitem{dulieu2011}
  O. Dulieu, R. Krems, M. Weidem\"uller, and S. Willitsch,
  Phys. Chem. Chem. Phys. {\bf 13}, 18703 (2011).
\bibitem{mol-papers}
  D. S. Jin and J. Ye,
  Chem. Rev. {\bf 112}, 4801 (2012), and references therein.


\bibitem{Mol-RC}
R. C\^ot\'e and A. Dalgarno,
J. Mol. Spectrosc. \textbf{195}, 236 (1999).

\bibitem{Pellegrini} 
E. Kuznetsova, M. Gacesa, P. Pellegrini, S. F. Yelin, and R. C\^ot\'e,
New J. Phys. \textbf{11}, 055028 (2009).


\bibitem{sawyer2011}
  B. C. Sawyer, B. K. Stuhl, M. Yeo, T. V. Tscherbul, M. T. Hummon, Y. Xia, 
  J. Klos, D. Patterson, J. M. Doyle, and J. Ye,
  Phys. Chem. Chem. Phys. {\bf 13}, 19059 (2011).
\bibitem{quemener2012}
  G. Qu\'em\'ener and P. Julienne,
  Chem. Rev. {\bf 112}, 4949 (2012).

\bibitem{DeMille-QC} D. DeMille, Phys. Rev. Lett. \textbf{88}, 067901 (2002).


\bibitem{Lena-1}
S. F. Yelin, K. Kirby, and R. C\^ot\'e,
Phys. Rev. A \textbf{74}, 050301 (2006).

\bibitem{Lena-2}
E. Kuznetsova, R. C\^ot\'e, K. Kirby, and S. F. Yelin,
Phys. Rev. A \textbf{78}, 012313 (2008).



\bibitem{Bohn:PRA87:2013}
  M. Mayle, G. Qu\'em\'ener, B. P. Ruzic, and J. L. Bohn,
  Phys. Rev. A {\bf 87}, 012709 (2013).

\bibitem{h2co-prl-2012}
  S. Chefdeville, T. Stoecklin, A. Bergeat,
  K. M. Hickson, C. Naulin, and  M. Costes,
  Phys. Rev. Lett. \textbf{109}, 023201 (2012).
\bibitem{shd-prl-2012}
  M. Lara, S. Chefdeville, K. M. Hickson, A. Bergeat, C. Naulin, J.-M. Launay, 
  and M. Costes,
  Phys. Rev. Lett. \textbf{109}, 133201 (2012).
\bibitem{narevicius-2012}
  A. B. Henson, S. Gersten, Y. Shagam, J. Narevicius, and E.~Narevicius,
  Science \textbf{338}, 234 (2012). 


\bibitem{alkali-hydrides}
N. Geum, G.-H. Jeung, A. Derevianko, R. C\^ot\'e, and A. Dalgarno,
J. Chem. Phys. \textbf{115}, 5984 (2001).


\bibitem{wigner}
  E. P. Wigner,
  Phys. Rev. \textbf{73}, 1002 (1948).

\bibitem{hossein}
H. R. Sadeghpour, J. L. Bohn, M. J. Cavagnero,
B. D. Esry, I. I. Fabrikant, J. H. Macek, and A. R. P. Rau,
J. Phys. B \textbf{33}, R93 (2000).

\bibitem{our-NTR} I. Simbotin, S. Ghosal, and R. C\^ot\'e,
  Phys. Rev. A, \textbf{89}, 040701 (2014).
\bibitem{Bethe:PR:1935} H. A. Bethe, Phys. Rev. \textbf{47}, 747 (1935).
\bibitem{fano} U. Fano, Phys. Rev. \textbf{124}, 1866 (1961).
\bibitem{alex-bala} N. Balakrishnan, R. C. Forrey, and A. Dalgarno,
  Phys. Rev. Lett. \textbf{80}, 3224 (1998).
\bibitem{bala-cplett} N. Balakrishnan, V. Kharchenko, R. C. Forrey,
  and A. Dalgarno, Chem. Phys. Lett. \textbf{280}, 5 (1997).
\bibitem{ClH2:Science:2008} E. Garand, J. Zhou, D. E. Manolopoulos,
  M. H. Alexander, and D. M. Neumark, Science \textbf{319}, 72 (2008).
\bibitem{Bala:H2Cl:2012} N. Balakrishnan, J. Chem. Sci. {\bf 124}, 311 (2012).
\bibitem{BWpes:JCP:2000} W. Bian and H.-J. Werner,
  J. Chem. Phys. \textbf{112}, 220 (2000).
\bibitem{H2F-Bala} N. Balakrishnan and A. Dalgarno,
  Chem. Phys. Lett. \textbf{341}, 652 (2001).
\bibitem{Bodo:JPB:2004} E. Bodo, F. A. Gianturco, N. Balakrishnan, and
  A. Dalgarno, J.~Phys. B \textbf{37}, 3641 (2004).
\bibitem{SWpes} K. Stark and H.-J. Werner, J. Chem. Phys. \textbf{104},
  6515 (1996).  
\bibitem{JMH:JCP:2007} M. T. Cvitas, P. Soldan, J. M. Hutson,
  P. Honvault, and \mbox{J.-M. Launay}, J. Chem. Phys. \textbf{127},
  074302 (2007).
\bibitem{ABC:CPC:2000} D. Skouteris, J. F. Castillo, and
  D. E. Manolopoulos, Comp. Phys. Comm. \textbf{133}, 128 (2000).
\bibitem{PCCP-H2+D} I. Simbotin, S. Ghosal, and R. C\^ot\'e,
  Phys. Chem. Chem. Phys. \textbf{13}, 19148 (2011).
\bibitem{NJPH2+D} I. Simbotin and R. C\^ot\'e, New J. Phys., in press (2015).
\bibitem{PRLH2+D} J. Wang, J. N. Byrd, I. Simbotin, and R. C\^ot\'e,
  Phys. Rev. Lett. \textbf{113}, 025302 (2014).
\bibitem{taylor} J. R. Taylor, \emph{Scattering Theory} (Dover
  Publications, Mineola, NY, 2006).
\bibitem{newton:0} See page 523 in Ref.~\cite{newton}.
\bibitem{newton} R. G. Newton, \emph{Scattering Theory of waves and
  particles} (Dover Publications, Mineola, NY, 2002).
\bibitem{newton:phi-psi} See page 343 in Ref.~\cite{newton}.
\bibitem{newton:page-335} See page 335 in Ref.~\cite{newton}.
\bibitem{gianturco-jost} K. Willner and F.  A. Gianturco, Phys. Rev. A
  \textbf{74}, 052715 (2006).
\bibitem{newton:page:xxx} See page 528 in Ref.~\cite{newton}.

\bibitem{rick.heller} E. J. Heller and W. P. Reinhardt, Phys. Rev. A
  \textbf{5}, 757 (1972).



\bibitem{Alex:PRA2010:He-H2}
  J. L. Nolte, B. H. Yang, P. C. Stancil, T.-G. Lee, N. Balakrishnan,
  R. C. Forrey,  and A. Dalgarno,
  Phys. Rev. A {\bf 81}, 014701 (2010).
 
\bibitem{JMH:PRL:2009:Feshbach}
  J. M. Hutson, M. Beyene, and M. L. Gonz\'alez-Mart\'\i{}nez, 
  Phys. Rev. Lett. \textbf{103}, 163201 (2009). 
\bibitem{NJP-Hutson-2007} J. M. Hutson, New J. Phys. \textbf{9}, 152 (2007).

\bibitem{Crubellier:JPB:2006}
  A. Crubellier and E. Luc-Koenig,
  J. Phys. B \textbf{39}, 1417 (2006).

   
\bibitem{FOPA}
  P. Pellegrini, M. Gacesa, and R. C\^ot\'e, 
  Phys. Rev. Lett. {\bf 101}, 053201 (2008).
\bibitem{Cs2:resonance}
  M. Arndt, M. Ben Dahan, D. Gu\'ery-Odelin, M. W. Reynolds, and J. Dalibard, 
  Phys. Rev. Lett. {\bf 79}, 625 (1997). 
   
\bibitem{Cs2:spin}
  J. S\"oding, D. Gu\'ery-Odelin, P. Desbiolles, G. Ferrari, and J.~Dalibard, 
  Phys. Rev. Lett. {\bf 80}, 1869 (1998).

\end{thebibliography}
\end{document}